\documentclass[10pt,journal]{IEEEtran}

\usepackage{amssymb}
\usepackage{amsmath}
\usepackage{cite}
\usepackage{url}
\usepackage{xcolor}
\usepackage{cite,graphicx,amsmath,amssymb}
\usepackage{fancyhdr}
\usepackage{mdwmath}
\usepackage{mdwtab}
\usepackage{caption}
\usepackage{amsthm}
\usepackage{setspace}
\usepackage{hyperref}
\usepackage{algorithm}
\usepackage{algorithmic}
\usepackage{multirow}
\usepackage{makecell}
\usepackage{mathtools}
\usepackage{subcaption}
\usepackage{mathrsfs}
\usepackage{bm}
\usepackage{pifont}
\usepackage{balance}

\usepackage{booktabs}
\usepackage{multirow}
\usepackage{siunitx}

\hypersetup{colorlinks=true,
linkcolor=blue,
citecolor=blue,      
urlcolor=black,
}

\graphicspath{{./src/img/}}

\DeclareMathOperator*{\argmax}{arg\,max}
\DeclareMathOperator*{\argmin}{arg\,min}
\newtheorem{remark}{Remark}
\newtheorem{theorem}{Theorem}

\newtheorem{lemma}{Lemma}

\newtheorem{corollary}{Corollary}

\allowdisplaybreaks
\title{Optimal Beamforming for Multi-User Continuous Aperture Array (CAPA) Systems}

\author{
        Zhaolin Wang,~\IEEEmembership{Member,~IEEE,}
        Chongjun Ouyang,~\IEEEmembership{Member,~IEEE,} \\
        and Yuanwei Liu,~\IEEEmembership{Fellow,~IEEE}
\thanks{Zhaolin Wang is with the School of Electronic Engineering and Computer Science, Queen Mary University of London, London E1 4NS, U.K. (e-mail: zhaolin.wang@qmul.ac.uk).}
\thanks{Chongjun Ouyang is with the School of Electrical and Electronic Engineering, University College Dublin, D04 V1W8, Ireland, and also with the School of Electronic Engineering and Computer Science, Queen Mary University of London, London, E1 4NS, U.K. (e-mail: chongjun.ouyang@ucd.ie).}
\thanks{Yuanwei Liu is with the Department of Electrical and Electronic Engineering, The University of Hong Kong, Hong Kong (e-mail: yuanwei@hku.hk).}
\vspace{-0.7cm}
}

\begin{document}

\maketitle
\begin{abstract}

    The optimal beamforming design for multi-user continuous aperture array (CAPA) systems is proposed. In contrast to conventional spatially discrete array (SPDA), the beamformer for CAPA is a continuous function rather than a discrete vector or matrix, rendering beamforming optimization a non-convex integral-based functional programming. To address this challenging issue, the closed-form optimal structure of the CAPA beamformer is first derived for maximizing generic system utility functions, by addressing the inversion of continuous functions and using the Lagrangian duality and the calculus of variations. The derived optimal structure is a linear combination of the continuous channel responses for CAPA, with the linear weights determined by the channel correlations. As a further advance, a monotonic optimization method is proposed for obtaining globally optimal CAPA beamforming based on the derived optimal structure. More particularly, a closed-form fixed-point iteration is proposed to obtain the globally optimal solution to the power minimization problem for CAPA beamforming. Furthermore, based on the optimal structure, the low-complexity maximum ratio transmission (MRT), zero-forcing (ZF), and minimum mean-squared error (MMSE) designs for CAPA beamforming are derived. It is theoretically proved that: 1) the MRT and ZF designs are asymptotically optimal in low and high signal-to-noise ratio (SNR) regimes, respectively, and 2) the MMSE design is optimal for signal-to-leakage-plus-noise ratio (SLNR) maximization. Our numerical results validate the effectiveness of the proposed designs and reveal that: \emph{i)} CAPA achieves significant communication performance gain over SPDA, and \emph{ii)} the MMSE design achieves nearly optimal performance in most cases, while the MRT and ZF designs achieve nearly optimal performance in specific cases.
    
\end{abstract}


\vspace{-0.3cm}
\section{Introduction} \label{sec:intro}

\IEEEPARstart{I}N the evolution of wireless communication, continuous aperture array (CAPA) has emerged as a transformative concept poised to redefine the boundaries of spatial multiplexing and capacity in wireless systems \cite{10220205, 10639537}. While CAPA has garnered significant attention in recent years, it can be traced back to the 1960s \cite{1138456, 1049891}, and has been a long-standing goal in antenna design. By building upon the foundational principles of multiple-input multiple-output (MIMO) technology, CAPA represents a leap forward in antenna architecture, enabling nearly continuous electromagnetic (EM) aperture coverage. Therefore, CAPA has the potential to break through the inherent physical constraints of conventional discrete antenna deployments, which limit the spatial degrees of freedom (DoFs) and ultimately the achievable spectral efficiency. As the wireless communication landscape transitions from Massive MIMO in 5G \cite{andrews2014will} to the envisioned Gigantic MIMO of 6G \cite{bjornson2024enabling}, CAPA stands out as a cornerstone concept. It harmonizes cutting-edge hardware innovations such as holographic MIMO \cite{ huang2020holographic, 9110848, 10232975}, large intelligent surfaces \cite{8319526, dardari2020communicating}, and dynamic metasurfaces \cite{shlezinger2021dynamic} into a unified framework, advancing the ultimate objective of achieving a continuous EM aperture. This makes CAPA not just a progression of antenna array technology but a pivotal enabler of future wireless communication paradigms.

\vspace{-0.3cm}
\subsection{Prior Works}

CAPA technique has attracted significant attention in wireless communication due to its potential to achieve unprecedented spatial DoFs and channel capacity. The unique continuous EM radiating surface of CAPA introduces both challenges and opportunities, driving extensive research in modeling, performance analysis, and system optimization.

The theoretical modeling of CAPA systems often employs EM theory, which represents transmit signals as sinusoidal source currents distributed across the aperture, generating propagating EM waves toward receivers. This approach, which connects information theory with EM theory, has formed the basis for analyzing CAPA performance. Research on the fundamental DoFs of CAPA systems dates back to the 1980s. Early work by \cite{bucci1989degrees} explored the DoFs of scattered EM fields based on Nyquist number over the observation interval. Later, the authors of \cite{miller2000communicating} formalized the relationship between DoFs and the physical volumes of CAPA transmitters and receivers through eigenfunction analysis. Expanding on this, the authors of \cite{dardari2020communicating} provided accurate analytical expressions for communication DoFs, demonstrating that even in line-of-sight (LoS) scenarios, CAPA systems achieve DoFs greater than one. Further advancements extended these insights to near-field scenarios. For instance, the authors of \cite{decarli2021communication} proposed a multi-focusing method to approximate optimal DoFs for large CAPAs. In \cite{9848802}, a spatial bandwidth approach was developed to derive closed-form DoF approximations for linear CAPAs in three orthogonal directions, followed by an investigation of effective DoFs in \cite{10262267}.

Channel capacity is another critical area of CAPA research. Using an eigenfunction approach, the authors of \cite{jensen2008capacity} established a general EM framework for CAPA capacity analysis, highlighting how finite transmit power imposes capacity bounds while offering practical solutions to mitigate supergain effects. Subsequently, the authors of \cite{jeon2017capacity} examined the effects of physical losses on both DoFs and capacity. Several studies have proposed innovative approaches to enhance capacity analysis and transmission strategies. For example, the authors of \cite{8585146} applied Kolmogorov information theory to evaluate CAPA capacity. In contrast, the authors of \cite{10303285} compared CAPA and conventional discrete MIMO systems, demonstrating that MIMO systems could approach CAPA performance by deploying an infinite number of antennas without mutual coupling. Recent advances include a numerical calculation scheme for random field capacity analysis proposed by \cite{wan2023mutual} and a general capacity computation method for arbitrary CAPA surfaces developed in \cite{10012689}. To enhance practical transmission efficiency, \cite{9906802} introduced a wavenumber-division multiplexing scheme, achieving near-optimal capacity under LoS conditions. By further utilizing the wavenumber-domain channel response, the authors of \cite{ouyang2024diversity} analyzed the ergodic channel capacity and diversity-multiplexing trade-off in CAPA-based fading channels.

While early studies primarily focused on single-transceiver CAPA systems, more recent research has addressed multi-user scenarios. For uplink systems under LoS conditions, the authors of \cite{hu2018beyond} analyzed uplink capacity using matched-filtering techniques. Building on this, the authors of \cite{10669060} investigated signal-to-interference-plus-noise ratio (SINR) optimization, proposing an adaptive interference mitigation method to enhance performance. The authors of \cite{zhao2024continuous} examined multi-user uplink and downlink capacities, utilizing uplink-downlink duality principles. From an optimization perspective, \cite{zhang2023pattern} introduced a Fourier-based beamforming method to maximize downlink sum rates, demonstrating significant performance gains for CAPA over conventional MIMO systems. This optimization approach was later adapted to uplink beamforming in multi-user systems by \cite{10612761}, reinforcing CAPA’s advantages in both transmission directions. To address the issue high computational complexity of the Fourier-based method, the authors of \cite{wang2024beamforming} proposed a more efficient and effective method for CAPA beamforming optimization based on the calculus of variations. Most recently, a deep learning method was proposed in \cite{guo2024multi} for CAPA beamforming. 

\vspace{-0.3cm}
\subsection{Motivation and Contributions}

The optimal beamforming solution has been a relentless pursuit for multi-antenna wireless communication systems \cite{5447076}. While obtaining the optimal beamforming solution can be computationally expensive, it serves as both a benchmark for evaluating system performance and a baseline for practical beamforming design. Moreover, the optimal beamforming solution often provides valuable insights into how system performance is influenced by various parameters. These insights have inspired the development of low-complexity yet highly efficient beamforming designs, such as maximum ratio transmission (MRT), zero-forcing (ZF), and minimum mean-squared error (MMSE) design \cite{6832894}, which are widely used in conventional spatially discrete array (SPDA) systems. 

\textcolor{black}{As discussed earlier, several optimization methods have been proposed for CAPA multi-user beamforming, including the Fourier-based method \cite{zhao2024continuous, 10612761}, the calculus of variations method \cite{wang2024beamforming}, and the deep learning method \cite{guo2024multi}. However, achieving a globally optimal design for CAPA multi-user beamforming remains challenging with these methods. Specifically, the Fourier-based method leverages a finite number of Fourier basis functions to approximate the continuous CAPA beamformers \cite{zhao2024continuous, 10612761}, which limits its ability to achieve truly optimal beamformers. While the calculus of variations method \cite{wang2024beamforming} and the deep learning method \cite{guo2024multi} avoid the finite and discrete approximations inherent in the Fourier-based approach, they are still not global optimization methods for CAPA beamforming. This is because they rely on suboptimal block coordinate descent or black-box neural networks, respectively. To the best of our knowledge, no existing works have addressed the global optimization of CAPA multi-user beamforming.} Additionally, the mathematical formulations of the MRT, ZF, and MMSE designs for CAPA systems, as well as their relationship to the optimal beamforming solution, remain unexplored. Motivated by these observations, we aim to study the optimal multi-user beamforming for CAPA systems. The main contributions of this paper are summarized as follows:
\begin{itemize}
    \item \textcolor{black}{We derive the closed-form structure of the globally optimal CAPA beamformers for maximizing general system utilities by addressing the inversion of continuous functions and leveraging the Lagrangian duality and the calculus of variations.} Our findings demonstrate that the optimal CAPA beamformer is a linear combination of the continuous channel responses, with the coefficients of the linear combination determined by the channel correlations. 
    \item \textcolor{black}{Based on this derived optimal structure, we propose a monotonic optimization method to achieve the globally optimal CAPA beamformers, leveraging the polyblock outer approximation algorithm. Furthermore, we propose a globally optimal fixed-point iteration approach to solve the transmit power minimization problem, aiming to find the projection of the polyblock vertex. This approach effectively transforms continuous integral-based calculations into simpler discrete matrix-based calculations.} 
    \item We derive the mathematical expressions of the low-complexity heuristic MRT, ZF, and MMSE designs for CAPA beamforming. Our findings indicate that the MRT and ZF designs are asymptotically optimal in the low and high signal-to-noise ratio (SNR) regimes, respectively. Furthermore, we theoretically prove that the MMSE design is optimal for maximizing the signal-to-leakage-and-noise ratio (SLNR) by again employing Lagrangian duality and the calculus of variations.
    \item We provide comprehensive numerical results to validate the effectiveness of the proposed designs. The results reveal a significant performance gain achieved by CAPAs compared to SPDAs in terms of communication sum rate. Moreover, the MMSE design achieves near-optimal performance in most cases, while the ZF and MRT designs achieve near-optimal performance in specific scenarios.
\end{itemize}

\subsection{Organization and Notations}
The structure of this paper is outlined as follows. Section \ref{sec:model} introduces the system model. The derivation of the optimal beamforming structure is detailed in Section \ref{sec:optimal_structure}, followed by the presentation of a global optimization method in Section \ref{sec:optimal_design}. In Section \ref{sec:Heuristic}, a low-complexity heuristic approach for CAPA beamforming is developed, including methods for MRT, ZF, and MMSE designs. Numerical evaluations and performance comparisons under various system configurations are presented in Section \ref{sec:results}. Finally, Section \ref{sec:conclusion} summarizes the findings and concludes the paper.

\emph{Notations:} Scalars are denoted using regular typeface, vectors and matrices are represented in boldface, and Euclidean subspaces are indicated with calligraphic letters. The set of complex, real numbers, and non-negative real numbers are denoted by $\mathbb{C}$, $\mathbb{R}$, and $\mathbb{R}_+$, respectively. The inverse, conjugate, transpose, conjugate transpose, and trace operators are denoted by $(\cdot)^{-1}$, $(\cdot)^*$, $(\cdot)^T$, $(\cdot)^H$, and $\mathrm{tr}(\cdot)$, respectively. The absolute value and Euclidean norm are denoted by $|\cdot|$ and $\|\cdot\|$. The ceiling operator is denoted by $\lceil \cdot \rceil$. The real part of a complex number of demoted by $\Re \{\cdot\}$. An identity matrix of dimension $N \times N$ is denoted by $\mathbf{I}_N$. The Dirac delta function on the space $\mathbb{R}^{N \times 1}$ is denoted by 
\begin{equation}
    \begin{aligned} \label{identify_function}
        \delta(\mathbf{s} - \mathbf{z}) &= 0 \quad \text{ for } \mathbf{s} \neq \mathbf{z}, \\
        \int_{\mathcal{V}} \delta(\mathbf{s} - \mathbf{z}) d \mathbf{s} &= 1,
    \end{aligned}
\end{equation}
where $\mathbf{s} \in \mathbb{R}^{N \times 1}$ and $\mathbf{z} \in \mathbb{R}^{N \times 1}$, and $\mathcal{V} \subseteq \mathbb{R}^{N \times 1}$ is any volume that contains the point $\mathbf{s} = \mathbf{z}$.
\section{System Model and Problem Formulation} \label{sec:model}

\begin{figure}[t!]
    \centering
    \includegraphics[width=0.4\textwidth]{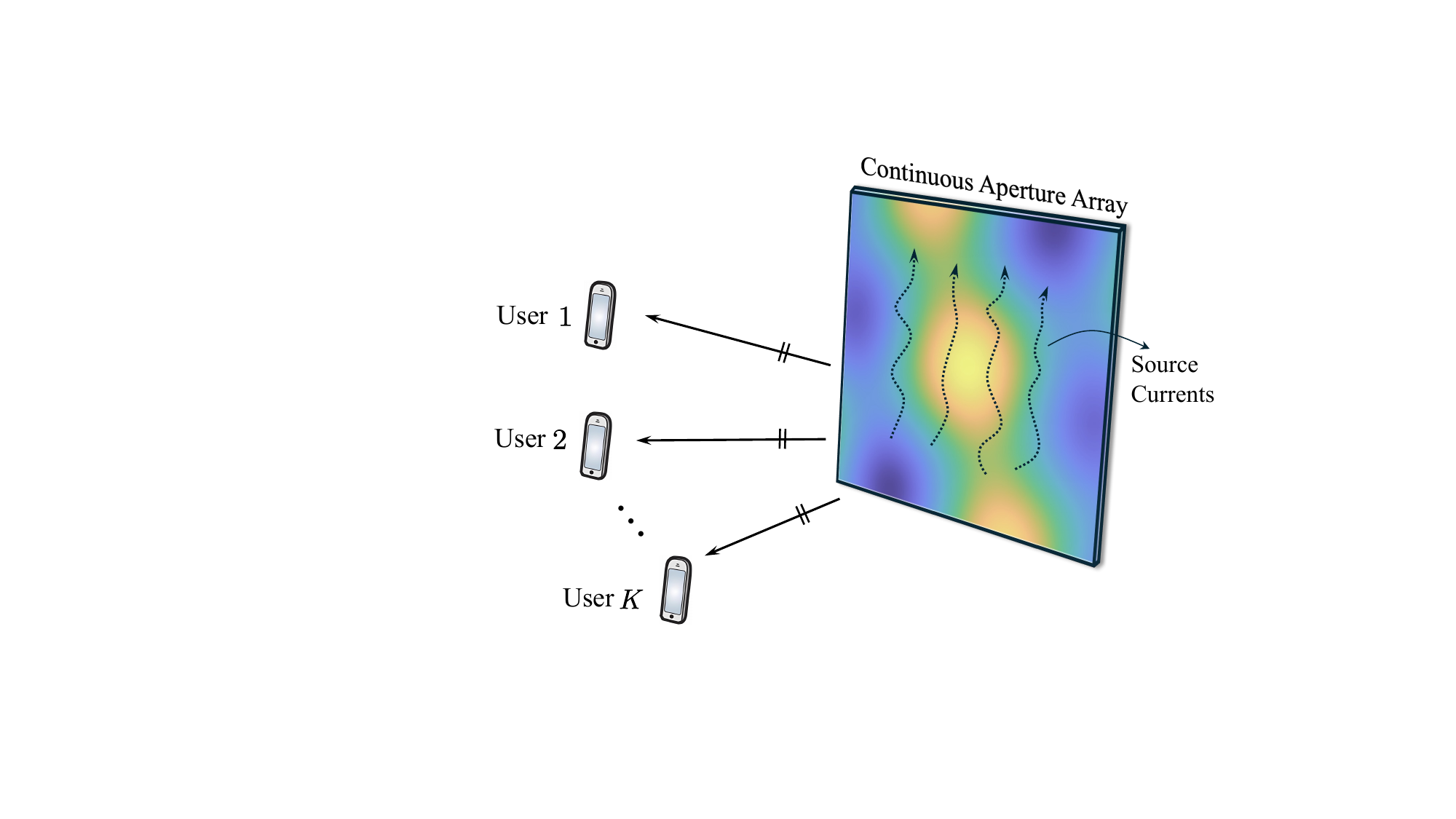}
    \caption{Illustration of CAPA-based multi-user communications.}
    \label{fig:system_model}
  \end{figure} 

As illustrated in Fig. \ref{fig:system_model}, we study a downlink multi-user communication system in which a uni-polarized CAPA transmitter at the base station serves $K$ communication users, each equipped with a single uni-polarized receive antenna. The CAPA transmitter features a continuous radiation surface $\mathcal{S}$, which carries sinusoidal source currents to emit electromagnetic waves for information transmission to the users. In this paper, we focus on a narrowband single-carrier communication system, where the density of the source current, i.e., the transmit signal, at a point $\mathbf{s} \in \mathcal{S}$ is modeled as
\begin{equation}
    x(\mathbf{s}) = \sum_{k=1}^K w_k(\mathbf{s}) c_k,
\end{equation}
where $w_k(\mathbf{s}) \in \mathbb{C}$ (\textcolor{black}{in amperes per meter, [$\mathrm{A/m}$]}) is the continuous beamformer representing the source current pattern across $\mathcal{S}$, used to convey the information symbol $c_k \in \mathbb{C}$ to user $k$. The information symbols are assumed to be independent and have unit average power, satisfying $\mathbb{E}\{ \mathbf{c} \mathbf{c}^H \} = \mathbf{I}_K$, where $\mathbf{c} = [c_1,\dots,c_K]^T$. The signal received at user $k$ is given by   
\vspace{-0.2cm} 
\begin{align} \label{receive_signal}
    y_k = &\int_{\mathcal{S}} h_k^*(\mathbf{s}) x(\mathbf{s}) d \mathbf{s} + n_k \nonumber \\
    = &\sum_{i=1}^K \int_{\mathcal{S}} h_k^*(\mathbf{s}) w_i(\mathbf{s}) c_i d \mathbf{s} + n_k,
\end{align}
where $h_k(\mathbf{s}) \in \mathbb{C}$ (\textcolor{black}{in ohms per square meter, [$\mathrm{\Omega/m^2}$]}) denotes channel response between the radiation point $\mathbf{s}$ and user $k$ and $n_k \sim \mathcal{CN}(0, \sigma^2)$ (\textcolor{black}{in volts per meter, [$\mathrm{V/m}$]}) denotes independent white Gaussian noise. The channel response $h_k(\mathbf{s})$ can be modelled to be either deterministic based on Green's function or stochastic based on random field theory. According to \eqref{receive_signal}, the signal-to-interference-plus-noise ratio (SINR) for decoding the desired information symbol at user $k$ is given by   
\begin{equation} \label{SINR}
    \gamma_k = \frac{\left| \int_{\mathcal{S}} h_k^*(\mathbf{s}) w_k(\mathbf{s}) d \mathbf{s} \right|^2}{\sum_{i\neq k} \left| \int_{\mathcal{S}} h_k^*(\mathbf{s}) w_i(\mathbf{s}) d \mathbf{s} \right|^2 + \sigma^2}.
\end{equation}

The communication performance can be evaluated using various metrics, most of which are monotonic functions of the SINR, including data rate, mean squared error, and bit error rate \cite{1223549}. Common system utility functions, such as the weighted arithmetic mean, weighted geometric mean, and max-min, are also monotonic functions of these metrics. Therefore, to explore the optimal structure of general CAPA beamforming, we consider a general optimization problem aimed at maximizing system performance with arbitrary monotonic utility functions, which is given by\footnote{\textcolor{black}{In practice, achieving a perfectly continuous radiation aperture can be challenging. However, various hardware designs have been developed to approximate a continuous radiation aperture, as discussed in \cite{liu2024capa}. Optimizing the transmit beamforming for these designs may introduce additional hardware-related constraints. As a result, the optimization problem in \eqref{problem_1} essentially serves as an upper-bound performance benchmark.}}
\begin{subequations} \label{problem_1}
    \begin{align}
       \max_{\mathbf{w}(\mathbf{s})} \quad &U(\boldsymbol{\gamma}) \\
       \label{power_constraint}
        \mathrm{s.t.} \quad & \sum_{k=1}^K \int_{\mathcal{S}} |w_k(\mathbf{s})|^2 d \mathbf{s} \le P,
    \end{align}
\end{subequations}
where $\mathbf{w}(\mathbf{s}) = [w_1(\mathbf{s}),\dots,w_K(\mathbf{s})] \in \mathbb{C}^{1 \times K}$ and $\boldsymbol{\gamma} = [\gamma_1,\dots,\gamma_K]^T \in \mathbb{R}^{K \times 1}$. Function $U(\cdot)$ is the system utility function that is strictly monotonically increasing in each argument $\gamma_k$. Constraint \eqref{power_constraint} is to ensure the transmit power does not exceed the budget $P$ (\textcolor{black}{in square amperes, [$\mathrm{A}^2$]})\footnote{\textcolor{black}{Although the power $P$ has a unit of $\mathrm{A}^2$, it is associated with two distinct types of power in unit of watt. \cite{Orfanidis, 9906802,zhang2023pattern}. The first type refers to the power delivered to the antenna terminals for generating the source currents $x(\mathbf{s})$. The second type represents the power of the EM field generated by the source current, which is computed by integrating the radial component of the Poynting vector over a sphere of radius $r \rightarrow +\infty$. These two power values are not necessarily identical and their relationship is determined by the antenna efficiency. To avoid ambiguity and maintain consistency in units, we use $\mathrm{A}^2$ to represent transmit power in the remaining part of this paper.}}. Although a similar problem to \eqref{problem_1} in SPDA systems has been extensively studied in the literature \cite{6832894}, solving problem \eqref{problem_1} remains challenging, as it is a functional programming problem where the optimal variables are no longer discrete vectors or matrices but continuous functions. In the following, we will investigate the structure of the optimal solution to problem \eqref{problem_1}, which will then be applied to achieve both globally optimal and efficient suboptimal designs.

\section{Optimal Structure of CAPA Beamforming} \label{sec:optimal_structure}

In this section, we will derive the structure of the optimal solution to problem \eqref{problem_1} as a function of the channel response for each user.

\subsection{Preliminary Lemmas}

Before delving into the details of deriving the optimal structure, we first present the following useful preliminary lemmas.

\begin{lemma} \label{Lemma_CoV}
    \normalfont
    \emph{(Fundamental Lemma of Calculus of Variations for Multi-variables)}    
    If a set of continuous functions $w_k(\mathbf{s}), k = 1,\dots,K,$ on $\mathcal{S}$ satisfies 
    \begin{equation} \label{fundamental_lemma_condition}
        \sum_{k=1}^K \Re \left\{ \int_{\mathcal{S}} w_k (\mathbf{s}) \eta_k^* (\mathbf{s}) d \mathbf{s} \right\} = 0,
    \end{equation}
    for any arbitrary smooth functions $\eta_k(\mathbf{s}), k = 1,\dots,K,$ defined on an open set $\mathcal{S}$ in the complex space, with the property that
    \begin{equation}
        \eta_k(\mathbf{s}) = 0, \forall \mathbf{s} \in \partial \mathcal{S},
    \end{equation}
    where $\partial \mathcal{S}$ is the boundary of $\mathcal{S}$, then it must follow that
    \begin{equation}
        w_k(\mathbf{s}) = 0, \forall \mathbf{s} \in \mathcal{S}.
    \end{equation}
\end{lemma}

\begin{IEEEproof}
    Please refer to Appendix \ref{Lemma_CoV_proof}
\end{IEEEproof}



\begin{lemma} \label{lemma_inverse}
    \normalfont \emph{(Inverse of Continuous Functions)} 
    Define the following function
    \begin{equation} \label{general_form}
        G(\mathbf{s}, \mathbf{z}) = \delta(\mathbf{s} - \mathbf{z}) + \sum_{i=1}^K \rho_i h_i(\mathbf{s}) h_i^*(\mathbf{z}).
    \end{equation}
    Its inverse is given by 
    \begin{equation} \label{general_form_inverse}
        G^{-1} (\mathbf{r}, \mathbf{s}) = \delta(\mathbf{r} - \mathbf{s}) - \sum_{k=1}^K \sum_{i=1}^K \rho_i d_{k,i} h_k(\mathbf{r}) h_i^*(\mathbf{s}),
    \end{equation}
    which satisfies 
    \begin{align}
        \label{left_inverse}
        & \int_{\mathcal{S}} G^{-1} (\mathbf{r}, \mathbf{s}) G(\mathbf{s}, \mathbf{z}) d \mathbf{s} = \delta(\mathbf{r} - \mathbf{z}).
    \end{align}
    Here, $d_{k,i}$ is the entry in the $k$-th row and $i$-th column of matrix $\mathbf{D} = (\mathbf{I}_K + \mathbf{\Lambda}\mathbf{Q})^{-1}$, where $\mathbf{\Lambda} = \mathrm{diag}\{ \rho_1,\dots,\rho_K \}$ is a diagonal matrix,
    and the entry of matrix $\mathbf{Q}$ in the $k$-th row and $i$-th column is given by 
    \begin{equation} \label{channel_correlation}
        q_{k,i} = [\mathbf{Q}]_{k,i} = \int_{\mathrm{S}} h_i(\mathbf{s}) h_k^*(\mathbf{s}) d \mathbf{s}.
    \end{equation}
\end{lemma}

\begin{IEEEproof}
    Please refer to Appendix \ref{lemma_inverse_proof}.
\end{IEEEproof}  

\begin{lemma} \label{lemma_power}
    \normalfont
    \emph{(Full Power Usage)} The optimal solution to problem \eqref{problem_1} satisfies the power constraint with equality, i.e.,
    \begin{equation}
        \sum_{k=1}^K \int_{\mathcal{S}} |w_k(\mathbf{s})|^2 d \mathbf{s} = P. 
    \end{equation}
\end{lemma}

\begin{IEEEproof}
    Please refer to Appendix \ref{lemma_power_proof}.
\end{IEEEproof}  

\subsection{Optimal Structure}

We are now ready to derive the optimal structure of CAPA beamforming. Inspired by \cite{6832894}, we approach this by utilizing the following power minimization problem:
\begin{subequations} \label{problem_2}
    \begin{align}
        \min_{\mathbf{w}(\mathbf{s})} \quad &\sum_{k=1}^K \int_{\mathcal{S}} |w_k(\mathbf{s})|^2 \, d\mathbf{s} \\
        \mathrm{s.t.} \quad & \gamma_k \ge \eta_k, \quad \forall k,
    \end{align}
\end{subequations}
where $\eta_k$ represents the minimum SINR required by user $k$. It can be proved that the optimal solutions to problems \eqref{problem_1} and \eqref{problem_2} share the same structure by substituting $\eta_k$ in problem \eqref{problem_2} with the optimal solution $\gamma_k^\star$ from problem \eqref{problem_1} \cite{6832894}. Then, we have the following theorem.

\begin{theorem} \label{theorem_structure}
    \normalfont
    \emph{(Optimal Structure of CAPA Beamforming)} The optimal solution to problems \eqref{problem_1} and \eqref{problem_2} can be expressed as 
    \begin{equation} \label{optimal_beamforming}
        \boxed{\mathbf{w}(\mathbf{s}) = \mathbf{h}(\mathbf{s}) \left(\mathbf{I}_K + \frac{1}{\sigma^2}\mathbf{\Lambda} \mathbf{Q} \right)^{-1} \mathbf{P}^{\frac{1}{2}},}
    \end{equation}
    where $\mathbf{h}(\mathbf{s}) = [h_1(\mathbf{s}), \dots, h_K(\mathbf{s})] \in \mathbb{C}^{1 \times K}$ is the channel vector, $\mathbf{Q} \in \mathbb{C}^{K \times K}$ is the channel correlation matrix with entries defined in \eqref{channel_correlation} satisfying $\mathbf{Q} = \mathbf{Q}^H$, and $\mathbf{\Lambda} = \mathrm{diag}(\lambda_1,\dots,\lambda_K) \in \mathbb{R}_+^{K \times K}$ and $\mathbf{P} = \mathrm{diag}(p_1,\dots,p_K) \in \mathbb{R}_+^{K \times K}$ are two diagonal matrices. Additionally, we have 
    \begin{equation} \label{lambda_condition}
        \sum_{k=1}^K \lambda_k = P,
    \end{equation}
    which implies that the sum of $\lambda$-parameters equals the maximum available transmit power. 
\end{theorem}

\begin{IEEEproof}
    Please refer to Appendix \ref{theorem_structure_proof}.
\end{IEEEproof}

\begin{remark}
    \normalfont
    The optimal structure \eqref{optimal_beamforming} comprises three components: the channel vector $\mathbf{h}(\mathbf{s})$, a rotation matrix $\left( \mathbf{I}_K + \mathbf{\Lambda} \mathbf{Q} / \sigma^2 \right)^{-1}$, and a power allocation matrix $\mathbf{P}$. The rotation matrix plays a critical role, as it determines the extent to which the beamformer for each user maximizes the power of the desired signal (reflected by the term $\mathbf{I}_K$) while mitigating interference to co-users (reflected by the term $\mathbf{\Lambda} \mathbf{Q}$).  
\end{remark}

\begin{remark} \label{remark_2}
    \normalfont
    The optimal structure \eqref{optimal_beamforming} indicates that the optimal CAPA beamformer for each user is a linear combination of the channel responses, given by $w_k(\mathbf{s}) = \sum_{i=1}^K a_{k,i} h_i(\mathbf{s})$, where $a_{k,i}$ is the entry in the $i$-th row and $k$-th column of the matrix $\left( \mathbf{I}_K + \mathbf{\Lambda} \mathbf{Q} / \sigma^2 \right)^{-1} \mathbf{P}^{\frac{1}{2}}$. In other words, the optimal CAPA beamformer resides within the subspace spanned by the channel responses. This result is intuitive and has a simple interpretation. Let $u_\perp(\mathbf{s})$ represent any function orthogonal to the subspace spanned by the channel responses, i.e., $\int_{\mathcal{S}} h_k^*(\mathbf{s}) u_\perp(\mathbf{s}) = 0, \forall k$. It can be shown that adding this component $u_\perp(\mathbf{s})$ to the beamformer $w_k(\mathbf{s})$ has no impact on the SINR of each user, implying that the optimal $w_k(\mathbf{s})$ should not include $u_\perp(\mathbf{s})$.
\end{remark}

\section{Globally Optimal Beamforming Design} \label{sec:optimal_design}

In this section, based on the derived optimal structure, we will introduce a globally optimal beamforming design for CAPA, using monotonic optimization techniques~\cite{zhang2013monotonic,bjornson2013optimal}.

\subsection{Problem Reformulation}

Problem \eqref{problem_1} can be transformed into a canonical monotonic optimization problem as 
\begin{subequations} \label{problem_4}
    \begin{align}
       \max_{\boldsymbol{\theta} } \quad &U(\boldsymbol{\theta}) \\
        \mathrm{s.t.} \quad & \boldsymbol{\theta} \in \mathcal{G},
    \end{align}
\end{subequations}
where $\boldsymbol{\theta} = [\theta_1,\dots,\theta_K]^T$ is a vector containing the new optimization variables and $\mathcal{G}$ is a normal set defined as 
\begin{equation}
    \mathcal{G} = \left\{ \boldsymbol{\theta}  \Big|  0 \le \theta_k \le \gamma_k, \sum_{k=1}^K \int_{\mathcal{S}} |w_k(\mathbf{s})|^2 d \mathbf{s} \le P \right\}.
\end{equation}
Although problem \eqref{problem_4} is non-convex, it has an important property that its optimum must lie on the Pareto boundary of the feasible region $\mathcal{G}$. This property allows us to achieve the optimum in polynomial time using techniques such as polyblock outer approximation, as detailed below.

\subsection{Polyblock Outer Approximation Algorithm}

According to \cite[Chapter 3]{zhang2013monotonic} and \cite[Chapter 2]{bjornson2013optimal}, the globally optimal $\boldsymbol{\theta}$ to problem \eqref{problem_4} can be obtained using the polyblock outer approximation algorithm summarized in \textbf{Algorithm \ref{alg:optimal}}. In this algorithm, the projection point $\pi_{\mathcal{G}}(\mathbf{z})$ and the initial point $\mathbf{b}$ are determined as follows.

\begin{algorithm}[tb]
    \caption{Polyblock Outer Approximation Algorithm for Obtaining the Globally Optimal Solution to Problem \eqref{problem_4}}
    \label{alg:optimal}
    \begin{algorithmic}[1]
        \STATE{Initialize the polyblock $\mathcal{P}_1$ as a box $[\mathbf{0}, \mathbf{b}]$ that enclosed the feasible set $\mathcal{G}$, the vertex set as $\mathcal{V}_1 = \{\mathbf{b}\}$, the solution accuracy $\varepsilon$, $U_{\min} = - \infty$, and $n = 1$}
        \REPEAT
        \STATE{Select the vertex $\mathbf{z}_n = \argmax \{U(\mathbf{z}) | \mathbf{z}\in \mathcal{V}_n\}$}
        \STATE{Set $U_{\max} = U(\mathbf{z}_n)$}
        \STATE{Calculate $\pi_{\mathcal{G}} (\mathbf{z}_n)$, i.e., the projection of $\mathbf{z}_n$ on the feasible set $\mathcal{G}$, according to \textbf{Algorithm \ref{alg:bisection}}}
        \IF{$ U(\pi_{\mathcal{G}} (\mathbf{z}_n)) > U_{\min}$}
        \STATE{set $\boldsymbol{\theta} =  \pi_{\mathcal{G}} (\mathbf{z}_n)$ and $U_{\min} = U(\pi_{\mathcal{G}} (\mathbf{z}_n))$}
        \ENDIF
        \STATE{Construct the new polyblock with vertices $\mathcal{V}_{n+1} = \left( \mathcal{V}_n \backslash \mathbf{z}_n \right) \cup \{ \tilde{\mathbf{z}}_k = \mathbf{z}_n - [\mathbf{z}_n - \pi_{\mathcal{G}} (\mathbf{z}_n)]_k \mathbf{e}_k | k = 1,\dots,K  \}$, where $\mathbf{e}_k$ denotes the $k$-th column of $\mathbf{I}_K$}
        \STATE{Set $n = n + 1$}
        \UNTIL{$|U_{\max} - U_{\min}| \le \varepsilon$ }
        \STATE{Output the solution $\boldsymbol{\theta}$}
    \end{algorithmic}
\end{algorithm}

\subsubsection{Finding the projection point $\pi_{\mathcal{G}} (\mathbf{z})$}

According to \cite[Proposition 3.3]{zhang2013monotonic}, the projection point can be expressed as 
\begin{equation}
    \pi_{\mathcal{G}} (\mathbf{z}) = \alpha \mathbf{z}, \quad \alpha = \max\{ \bar{\alpha} > 0 | \bar{\alpha} \mathbf{z} \in \mathcal{G} \}.
\end{equation}
Hence, determining the projection point $\pi_{\mathcal{G}} (\mathbf{z})$ reduces to finding the scalar $\alpha$, which can be efficiently obtained through a bisection search. For any given $\alpha > 0$, the feasibility condition $\alpha \mathbf{z} \in \mathcal{G}$ can be verified by solving the following optimization problem:
\begin{subequations} \label{problem_feasibility_check}
    \begin{align}
       \min_{ \mathbf{w}(\mathbf{s}) } \quad &\sum_{k=1}^K \int_{\mathcal{S}} |w_k(\mathbf{s})|^2 d \mathbf{s} \\
        \mathrm{s.t.} \quad & \gamma_k  \ge \alpha z_k,
    \end{align}
\end{subequations}
where $z_k$ denotes the $k$-th entry of $\mathbf{z}$. Let $P^{\star}$ denote the optimal objective value of the above problem. Then, we have if $P^{\star} \le P$, $\alpha \mathbf{z} \in \mathcal{G}$; otherwise, $\alpha \mathbf{z} \notin \mathcal{G}$.
Based on \eqref{dual_gap} in Appendix \ref{theorem_structure_proof}, it can be shown that 
\begin{equation} \label{feasibility_P}
    P^{\star} = \sum_{k=1}^K \lambda_k,
\end{equation}
where $\boldsymbol{\lambda} = [\lambda_1,\dots,\lambda_K]^T$ are the Lagrange multiplier of problem \eqref{problem_feasibility_check}. The optimal $\lambda_k$ can be obtained by using the optimal structure \eqref{scalar_optimal_structure} (i.e., the original expression of the optimal structure in \textbf{Theorem \ref{theorem_structure}}), which implies 
\begin{align} \label{optimal_structure_lambda}
    w_k(\mathbf{s}) = &\frac{\lambda_k}{\sigma^2} \left(1 + \frac{1}{\alpha z_k}\right) \int_{\mathcal{S}} h_k^*(\mathbf{z}) w_k(\mathbf{z}) d \mathbf{z} \nonumber\\
    &\times \left(h_k(\mathbf{s}) - \sum_{i=1}^K \sum_{j=1}^K \frac{\lambda_j}{\sigma^2} d_{i,j} q_{j,k} h_i(\mathbf{s}) \right),
\end{align}
where $d_{i,j}$ is the entry of matrix $\mathbf{D} = (\mathbf{I}_K + \frac{1}{\sigma^2}\mathbf{\Lambda}\mathbf{Q})^{-1}$ in the $i$-th row and $j$-th column, with $\mathbf{\Lambda} = \mathrm{diag} \left( \lambda_1,\dots,\lambda_K \right)$, and $q_{j,k}$ is the channel correlation defined in \eqref{channel_correlation}. Multiplying both sides of \eqref{optimal_structure_lambda} by $h_k^*(\mathbf{s})$ and integrating over $d \mathbf{s}$, we have 
\begin{align} \label{optimal_structure_lambda_2}
    &\hspace{-0.8cm} \int_{\mathcal{S}} h_k^*(\mathbf{s})  w_k(\mathbf{s}) d \mathbf{s} \nonumber\\ = & \frac{\lambda_k}{\sigma^2} \left(1 + \frac{1}{\alpha z_k}\right) \int_{\mathcal{S}} h_k^*(\mathbf{z}) w_k(\mathbf{z}) d \mathbf{z} \nonumber\\
    & \times \int_{\mathcal{S}} h_k^*(\mathbf{s}) \left(  h_k(\mathbf{s})  - \sum_{i=1}^K \sum_{j=1}^K \frac{\lambda_j}{\sigma^2} d_{i,j} q_{j,k} h_i(\mathbf{s}) \right) d \mathbf{s} \nonumber \\
    = & \frac{\lambda_k}{\sigma^2} \left(1 + \frac{1}{\alpha z_k}\right) \int_{\mathcal{S}} h_k^*(\mathbf{z}) w_k(\mathbf{z}) d \mathbf{z} \nonumber \\
    &\times \left(  q_{k,k}  - \sum_{i=1}^K \sum_{j=1}^K \frac{\lambda_j}{\sigma^2} d_{i,j} q_{j,k} q_{k,i} \right).
\end{align} 
Then, dividing both sides of \eqref{optimal_structure_lambda_2} by $\int_{\mathcal{S}} h_k^*(\mathbf{s})  w_k(\mathbf{s}) d \mathbf{s}$ yields
\begin{align} \label{scalar_optimal_lambda}
    1 = & \frac{\lambda_k}{\sigma^2} \left(1 + \frac{1}{\alpha z_k}\right) \left(  q_{k,k}  - \sum_{i=1}^K \sum_{j=1}^K \frac{\lambda_j}{\sigma^2} d_{i,j} q_{j,k} q_{k,i} \right) \nonumber \\
    = & \frac{\lambda_k}{\sigma^2} \left(1 + \frac{1}{\alpha z_k}\right) \left( q_{k,k} - \frac{1}{\sigma^2} \bar{\mathbf{q}}_k^T  \mathbf{D} \mathbf{\Lambda} \mathbf{q}_k \right),
\end{align} 
where $\mathbf{q}_k$ and $\bar{\mathbf{q}}_k$ are the vectors in the $k$-th column of $\mathbf{Q}$ and $\mathbf{Q}^T$, respectively. \eqref{scalar_optimal_lambda} can be written in matrix form as 
\begin{align} \label{matrix_optimal_lambda}
    \mathbf{I}_K = & \frac{1}{\sigma^2} \mathbf{\Lambda} \mathbf{Z}_{\alpha}  \mathrm{diag}\left( \mathbf{Q} - \frac{1}{\sigma^2} \mathbf{Q} \mathbf{D} \mathbf{\Lambda} \mathbf{Q} \right) \nonumber \\
    = & \frac{1}{\sigma^2} \mathbf{\Lambda} \mathbf{Z}_{\alpha}  \mathrm{diag}\left( \mathbf{Q} \left( \mathbf{I}_k -  \left(\mathbf{I}_K + \frac{1}{\sigma^2}\mathbf{\Lambda}\mathbf{Q}\right)^{-1} \frac{1}{\sigma^2} \mathbf{\Lambda} \mathbf{Q} \right) \right) \nonumber \\
    \overset{(a)}{=} & \frac{1}{\sigma^2} \mathbf{\Lambda} \mathbf{Z}_{\alpha}  \mathrm{diag}\left( \mathbf{Q} \left( \mathbf{I}_K + \frac{1}{\sigma^2} \mathbf{\Lambda} \mathbf{Q} \right)^{-1} \right),
\end{align}
where the last step is derived from the Woodbury matrix identity and $\mathbf{Z}_{\alpha} = \mathrm{diag}(1 + \frac{1}{\alpha z_1},\dots,1 + \frac{1}{\alpha z_K})$. Since both sides of \eqref{matrix_optimal_lambda} are diagonal matrices, we can obtain the following condition of the optimal $\boldsymbol{\lambda}$ from \eqref{matrix_optimal_lambda}:
\begin{align}
    \lambda_k = &\frac{\sigma^2}{\left(1 + \frac{1}{\alpha z_k}\right) \left[ \mathbf{Q} \left( \mathbf{I}_K + \frac{1}{\sigma^2} \mathbf{\Lambda} \mathbf{Q} \right)^{-1} \right]_{k,k}} \nonumber \\
    = & f_k(\alpha, \boldsymbol{\lambda}), \forall k.
\end{align}
Note that the optimal $\lambda_k$ is the fixed point of the function $f_k(\alpha, \boldsymbol{\lambda})$, which can be obtained by the following simple fixed-point iteration:
\begin{equation} \label{fixed_point}
    \lambda_k^{(n+1)} = f_k \left(\alpha, \boldsymbol{\lambda}^{(n)} \right), \forall k.
\end{equation}

Based on the above solutions, the bisection search process for finding the projection point $\pi_{\mathcal{G}} (\mathbf{z})$ is summarized in \textbf{Algorithm \ref{alg:bisection}}.

\subsubsection{Finding the initial point $\mathbf{b}$} The initial point $\mathbf{b}$ should be chosen so that the box $[\mathbf{0}, \mathbf{b}]$ fully encloses the feasible set $\mathcal{G}$. This requirement is equivalent to ensuring that $\mathbf{b} \geq \max \boldsymbol{\gamma}$. A closed-form upper bound of $\boldsymbol{\gamma}$ can be obtained as follows:
\begin{align}
    \gamma_k \overset{(a)}{\le} & \frac{\left| \int_{\mathcal{S}} h_k^*(\mathbf{s}) w_k(\mathbf{s}) d \mathbf{s} \right|^2}{\sigma^2} \nonumber \\ \overset{(b)}{\le} & \frac{P \left| \int_{\mathcal{S}} h_k^*(\mathbf{s}) h_k(\mathbf{s}) d \mathbf{s} \right|^2}{\sigma^2 \int_{\mathcal{S}} |h_k(\mathbf{s)}|^2 d \mathbf{s}}
    =  \frac{P q_{k,k}}{\sigma^2},
\end{align} 
where inequality $(a)$ is obtained by removing the interference terms in $\gamma_k$ and inequality $(b)$ is obtained by allocating all transmit power to user $k$ and aligning $w_k(\mathbf{s})$ with $h_k(\mathbf{s})$. Therefore, the $k$-th entry of $\mathbf{b}$, denoted by $b_k$ can be set as 
\begin{equation}
    b_k = \frac{P \left| \int_{\mathcal{S}} h_k^*(\mathbf{s}) h_k(\mathbf{s}) d \mathbf{s} \right|^2}{\sigma^2 \int_{\mathcal{S}} |h_k(\mathbf{s)}|^2 d \mathbf{s}}.
\end{equation}

After obtaining the globally optimal solution $\boldsymbol{\theta}^{\star} = [\theta_1^{\star}, \dots, \theta_K^{\star}]^T$ for problem \eqref{problem_4} via \textbf{Algorithm \ref{alg:optimal}}, the corresponding optimal beamformer $\mathbf{w}(\mathbf{s})$ can be derived by solving the following set of equations:
\begin{equation}
    \frac{\left| \int_{\mathcal{S}} h_k^*(\mathbf{s}) w_k(\mathbf{s}) \, d\mathbf{s} \right|^2}{\sum_{i \neq k} \left| \int_{\mathcal{S}} h_k^*(\mathbf{s}) w_i(\mathbf{s}) \, d\mathbf{s} \right|^2 + \sigma^2} = \theta_k^{\star}, \quad \forall k.
\end{equation} 
To solve these equations, one approach is to address problem \eqref{problem_feasibility_check} by substituting $\alpha z_k$ with $\theta_k^{\star}$, using the structure of the optimal solution in \eqref{optimal_structure_lambda} along with the fixed-point iteration described in \eqref{fixed_point}. The theoretical proof of the global optimality achieved by \textbf{Algorithm \ref{alg:optimal}} can be found in \cite[Chapter 3]{zhang2013monotonic} and \cite[Chapter 2]{bjornson2013optimal}.
Although \textbf{Algorithm \ref{alg:optimal}} is capable of finding the globally optimal solution for CAPA beamforming, it has a prohibitive complexity growing exponentially with the number of users $K$. Therefore, it is essential to develop low-complexity CAPA beamforming designs in practice.

\begin{algorithm}[tb]
    \caption{Bisection Search for Finding $\pi_{\mathcal{G}} (\mathbf{z})$}
    \label{alg:bisection}
    \begin{algorithmic}[1]
        \STATE{Initialize upper bound $\alpha_{\max} = 1$, lower bound $\alpha_{\min} = 0$, and the solution accuracy $\varepsilon$}
        \REPEAT
        \STATE{Select $\alpha = (\alpha_{\max} + \alpha_{\min})/2$  }
        \STATE{Calculate $\lambda_k, \forall k,$ using the fixed-point iteration \eqref{fixed_point}} 
        \STATE{Calculate $P^{\star} = \sum_{k=1}^K \lambda_k$}
        \IF{$P^{\star} \le P$ }
        \STATE{Update $\alpha_{\min} = \alpha$}
        \ELSE  
        \STATE{Update $\alpha_{\max} = \alpha$}
        \ENDIF
        \UNTIL{$\alpha_{\max} - \alpha_{\min} \le \varepsilon$ }
        \STATE{Calculate the projection point as $\pi_{\mathcal{G}}(\mathbf{z}) = \alpha \mathbf{z}$ }
    \end{algorithmic}
\end{algorithm}
\section{Low-Complexity Beamforming Design} \label{sec:Heuristic}
In this section, the low-complexity heuristic MRT, ZF, and MMSE beamforming designs for CAPA are developed based on the optimal structure derived in Section \ref{sec:optimal_structure}. 

\subsection{Asymptotically Optimal MRT and ZF Designs}

For conventional SPDAs, MRT and ZF beamforming designs are asymptotically optimal in low and high signal-to-noise (SNR) regimes, respectively. This result still holds for CAPAs, as elaborated below.

\subsubsection{MRT Design for Low-SNR Regime}

In the low-SNR regime, where $\sigma^2 \rightarrow \infty$, the optimal CAPA beamforming expression in \eqref{optimal_beamforming} simplifies to
\begin{equation}
    \mathbf{w}(\mathbf{s}) \xrightarrow{\sigma^2 \rightarrow \infty} \mathbf{h}(\mathbf{s}) (\mathbf{I}_K + \mathbf{0})^{-1} \mathbf{P}^{\frac{1}{2}} = \mathbf{h}(\mathbf{s}) \mathbf{P}^{\frac{1}{2}}.
\end{equation}
This indicates that the optimal beamformer for each user aligns with its respective channel, meaning that in the low SNR case, $w_k(\mathbf{s}) = \sqrt{p_k} h_k(\mathbf{s})$. By applying the Cauchy-Schwarz inequality,
\begin{equation} \label{Cauchy-Schwarz}
    \left|\int_{\mathcal{S}} h_k^*(\mathbf{s}) w_k(\mathbf{s}) d \mathbf{s}\right|^2 \le \int_{\mathcal{S}} \left|h_k(\mathbf{s})\right|^2 d \mathbf{s} \int_{\mathcal{S}} \left|w_k(\mathbf{s})\right|^2 d \mathbf{s},
\end{equation}
it becomes clear that this beamforming design is essentially the MRT design, which maximizes the received signal power $|\int_{\mathcal{S}} h_k^*(\mathbf{s}) w_k(\mathbf{s}) d \mathbf{s}|^2$ for each user, i.e., the equality in \eqref{Cauchy-Schwarz} is achieved. Thus, we define the MRT design as
\begin{equation}
    \boxed{\mathbf{w}_{\mathrm{MRT}}(\mathbf{s}) = \mathbf{h}(\mathbf{s})\mathbf{P}^{\frac{1}{2}}.}
\end{equation}

\subsubsection{ZF Design for High-SNR Regime} In the high-SNR regime, where $\sigma^2 \rightarrow 0$, the optimal CAPA beamforming in \eqref{optimal_beamforming} simplifies to
\begin{equation} \label{high_SNR}
    \mathbf{w}(\mathbf{s}) \xrightarrow{\sigma^2 \rightarrow 0} \mathbf{h}(\mathbf{s}) (0\mathbf{I}_K + \mathbf{\Lambda} \mathbf{Q})^{-1} \mathbf{P}^{\frac{1}{2}} = \mathbf{h}(\mathbf{s}) \mathbf{Q}^{-1} \tilde{\mathbf{P}}^{\frac{1}{2}},
\end{equation}
where $\tilde{\mathbf{P}} = \mathrm{diag}(p_1/\lambda_1^2, \dots, p_K/\lambda_K^2) \in \mathbb{R}_+^{K \times K}$. This beamforming design is essentially the ZF design, which completely eliminates inter-user interference. This can be demonstrated as follows:
\begin{align}
    &\begin{bmatrix}
         \int_{\mathcal{S}} h_1^*(\mathbf{s}) w_1(\mathbf{s})   d \mathbf{s} & \cdots & \int_{\mathcal{S}} h_1^*(\mathbf{s}) w_K(\mathbf{s})   d \mathbf{s} \\
         \vdots  & \ddots & \vdots \\
         \int_{\mathcal{S}} h_K^*(\mathbf{s}) w_1(\mathbf{s})   d \mathbf{s} & \cdots & \int_{\mathcal{S}} h_K^*(\mathbf{s}) w_K(\mathbf{s})   d \mathbf{s}
    \end{bmatrix} \nonumber \\
    &= \int_{\mathcal{S}} \mathbf{h}^H(\mathbf{s}) \mathbf{w}(\mathbf{s}) d \mathbf{s} \overset{(a)}{=} \int_{\mathcal{S}} \mathbf{h}^H(\mathbf{s}) \mathbf{h}(\mathbf{s}) \mathbf{Q}^{-1} \tilde{\mathbf{P}}^{\frac{1}{2}} d \mathbf{s} \nonumber \\
    &\overset{(b)}{=} \mathbf{Q} \mathbf{Q}^{-1} \tilde{\mathbf{P}}^{\frac{1}{2}} = \tilde{\mathbf{P}}^{\frac{1}{2}},
\end{align}
where step $(a)$ follows from substituting \eqref{high_SNR}, and step $(b)$ arises from the fact that $\mathbf{Q} = \int_{\mathcal{S}} \mathbf{h}^H(\mathbf{s}) \mathbf{h}(\mathbf{s})  d \mathbf{s}$. Since $\tilde{\mathbf{P}}$ is a diagonal matrix, it follows that $\int_{\mathcal{S}} h_k^*(\mathbf{s}) w_i(\mathbf{s})  d \mathbf{s} = 0$ for all $k \neq i$, which confirms that the inter-user interference has been eliminated. Therefore, we define the ZF design as
\begin{equation}
    \boxed{\mathbf{w}_{\mathrm{ZF}}(\mathbf{s}) = \mathbf{h}(\mathbf{s}) \mathbf{Q}^{-1} \mathbf{P}^{\frac{1}{2}}.}
\end{equation}
\textcolor{black}{Here, we replace $\tilde{\mathbf{P}}$ with $\mathbf{P}$ to maintain consistency with the other design. Note that this replacement does not affect the structure of the ZF design, as each entry of $\mathbf{P}$ is simply a linearly scaled version of the corresponding entries in $\tilde{\mathbf{P}}$.}

\subsection{MMSE Design}
Although the MRT and the ZF designs are optimal in the asymptotic cases with low and high SNR, respectively, they do not have the optimal structure \eqref{optimal_beamforming} in general cases. A common heuristic design for realizing the close-to-optimal beamforming is to select the $\lambda$-parameters in the optimal structure \eqref{optimal_beamforming} heuristically. One of the most straightforward ways is to set all $\lambda$-parameters to the same value, i.e., $\lambda_k = \tilde{\lambda}, \forall k$. According to \eqref{lambda_condition}, it can be shown that $\tilde{\lambda} = P/K$. Under this setup, the MMSE design (which is also commonly referred to as regularized ZF design) can be obtained, which is given by 
\begin{equation} \label{MMSE_deaign}
    \boxed{\mathbf{w}_{\mathrm{MMSE}} (\mathbf{s}) = \mathbf{h}(\mathbf{s}) \left(\mathbf{I}_K + \frac{P}{K \sigma^2} \mathbf{Q} \right)^{-1} \mathbf{P}^{\frac{1}{2}}.}
\end{equation} 

For conventional SPDAs, the effectiveness of the MMSE design lies in its ability to maximize the ratio of the desired signal power received at each user to the interference-plus-noise power caused to other users, known as the SLNR \cite{6832894}. This conclusion can be extended to CAPAs based on the following theorem.

\begin{theorem} \label{theorem_SLNR}
    \normalfont
    \emph{(SLNR-Optimal Design)} The MMSE design in \eqref{MMSE_deaign} maximizes the SLNR of each user in CAPA systems under the equal power allocation, i.e.,
    \begin{equation} \label{problem_SLNR_max}
        w_{\mathrm{MMSE}, k} (\mathbf{s}) = \argmax_{w_k(\mathbf{s}): \int_{\mathcal{S}} |w_k(\mathbf{s})|^2 d \mathbf{s} = \frac{P}{K}} \mathrm{SLNR}_k,
    \end{equation}
    where $w_{\mathrm{MMSE}, k} (\mathbf{s})$ is the $k$-th entry of $\mathbf{w}_{\mathrm{MMSE}} (\mathbf{s})$ with the power scaled to $P/K$ and $\mathrm{SLNR}_k$ is defined as       
    \begin{equation} \label{SLNR}
        \mathrm{SLNR}_k = \frac{\left| \int_{\mathcal{S}} h_k^*(\mathbf{s}) w_k(\mathbf{s}) d \mathbf{s} \right|^2}{\sum_{i\neq k} \left| \int_{\mathcal{S}} h_i^*(\mathbf{s}) w_k(\mathbf{s}) d \mathbf{s} \right|^2 + \sigma^2}.
    \end{equation}
\end{theorem}

\begin{IEEEproof}
    Please refer to Appendix \ref{theorem_SLNR_proof}
\end{IEEEproof}

\begin{table*}
    \centering
    \caption{Comparison of the Multi-User Beamforming Designs for CAPAs and SPDAs.}
    \label{table_comparison}
    \begin{tabular}{c!{\vrule width1pt}c!{\vrule width1pt}c}
        \Xhline{1pt}
        \textbf{Design method} & \textbf{CAPA} & \textbf{SPDA} \\
        \Xhline{1pt}
        Optimal Structure & $\mathbf{w}(\mathbf{s}) = \mathbf{h}(\mathbf{s}) \left(\mathbf{I}_K + \frac{1}{\sigma^2}\mathbf{\Lambda} \mathbf{Q} \right)^{-1} \mathbf{P}^{\frac{1}{2}}$ & $\mathbf{W} = \mathbf{H} \left( \mathbf{I}_K + \frac{1}{\sigma^2} \mathbf{\Lambda} \mathbf{H}^H \mathbf{H} \right)^{-1} \mathbf{P}^{\frac{1}{2}}$\\
        MRT Design & $\mathbf{w}_{\mathrm{MRT}}(\mathbf{s}) = \mathbf{h}(\mathbf{s})\mathbf{P}^{\frac{1}{2}}$ & $\mathbf{W}_{\mathrm{MRT}} = \mathbf{H} \mathbf{P}^{\frac{1}{2}}$ \\
        ZF Design &$\mathbf{w}_{\mathrm{ZF}}(\mathbf{s}) = \mathbf{h}(\mathbf{s}) \mathbf{Q}^{-1} \mathbf{P}^{\frac{1}{2}}$ & $\mathbf{W}_{\mathrm{ZF}} = \mathbf{H}\left( \mathbf{H}^H \mathbf{H} \right)^{-1} \mathbf{P}^{\frac{1}{2}}$  \\
        MMSE Design &$\mathbf{w}_{\mathrm{MMSE}} (\mathbf{s}) = \mathbf{h}(\mathbf{s}) \left(\mathbf{I}_K + \frac{P}{K \sigma^2} \mathbf{Q} \right)^{-1} \mathbf{P}^{\frac{1}{2}}$ & $\mathbf{W}_{\mathrm{MMSE}} = \mathbf{H} \left( \mathbf{I}_K + \frac{P}{K \sigma^2} \mathbf{H}^H \mathbf{H} \right)^{-1} \mathbf{P}^{\frac{1}{2}}$  \\
        \Xhline{1pt}
    \end{tabular}
\end{table*}

\subsection{Power Allocation} \label{sec:power_allocation}
Now, we turn our attention to designing the power allocation matrix $\mathbf{P}$. For the aforementioned MRT, ZF, and MMSE beamforming designs, the beamformer can be represented in the following general form:
\begin{equation}
    \mathbf{w}(\mathbf{s}) = \mathbf{h}(\mathbf{s}) \mathbf{\Phi} \mathbf{P}^{\frac{1}{2}},
\end{equation}
where $\mathbf{\Phi}$ is specified as $\mathbf{I}_K$ for the MRT design, $\mathbf{Q}^{-1}$ for the ZF design, and $\left(\mathbf{I}_K + \frac{P}{K \sigma^2} \mathbf{Q}\right)^{-1}$ for the MMSE design. As a consequence, the beamformer for user $k$ is given by 
\begin{equation}
    w_k(\mathbf{s}) = \sqrt{p_k} \mathbf{h}(\mathbf{s}) \boldsymbol{\phi}_k, 
\end{equation}
where $\boldsymbol{\phi}_k$ is the vector in the $k$-th column of $\mathbf{\Phi}$. Then, 
the SINR for user $k$ can be rewritten as  
\begin{align}
    \gamma_k = &\frac{ \left| \int_{\mathcal{S}} \sqrt{p_k} h_k^*(\mathbf{s}) \mathbf{h}(\mathbf{s}) \boldsymbol{\phi}_k d \mathbf{s} \right|^2}{\sum_{i\neq k} \left| \int_{\mathcal{S}} \sqrt{p_i} h_k^*(\mathbf{s}) \mathbf{h}(\mathbf{s}) \boldsymbol{\phi}_i d \mathbf{s} \right|^2 + \sigma^2} \nonumber \\
    = &\frac{p_k \left| \mathbf{q}_k^H \boldsymbol{\phi}_k \right|^2}{\sum_{i\neq k} p_i \left| \mathbf{q}_k^H \boldsymbol{\phi}_i \right|^2 + \sigma^2},
\end{align}
where $\mathbf{q}_k = \int_{\mathcal{S}} \mathbf{h}^H(\mathbf{s}) h_k(\mathbf{s}) d \mathbf{s}$ is the vector in the $k$-th column of $\mathbf{Q}$. Furthermore, the transmit power becomes 
\begin{align}
    \sum_{k=1}^K \int_{\mathcal{S}} |w_k(\mathbf{s})|^2 d \mathbf{s} = &\sum_{k=1}^K p_k \int_{\mathcal{S}} \boldsymbol{\phi}_k^H \mathbf{h}^H(\mathbf{s}) \mathbf{h}(\mathbf{s}) \boldsymbol{\phi}_k d \mathbf{s} \nonumber \\
    = &\sum_{k=1}^K p_k \boldsymbol{\phi}_k^H \mathbf{Q} \mathbf{\phi}_k.
\end{align}
The power allocation problem is thus formulated as
\begin{subequations} \label{problem_power}
    \begin{align}
       \max_{\mathbf{p}} \quad & U(\boldsymbol{\gamma}) \\
        \mathrm{s.t.} \quad & \sum_{k=1}^K p_k \boldsymbol{\phi}_k^H \mathbf{Q} \boldsymbol{\phi}_k \le P,
    \end{align}
\end{subequations}
where $\mathbf{p} = [p_1, \dots, p_K]^T$. This problem can be regarded as a power allocation problem for single-input single-output (SISO) systems. Different utility function $U(\cdot)$ results in a different solution. A comprehensive analysis of solutions for various $U(\cdot)$ is available in \cite[Section 2.2.4]{bjornson2013optimal} and \cite[Section 3.4.4]{bjornson2013optimal}. Hence, we omit the detailed discussion here.

\subsection{Comparison with SPDA Beamforming Designs} \label{section:SPDA}

To gain further insights, we compare the CAPA and SPDA beamforming designs. Let us consider an $N$-antenna SPDA transmitter serving $K$ single-antenna users. Under this setup, the SINR for decoding the desired signal at user $k$ can be expressed as 
\begin{equation} \label{SPDA_SINR}
    \tilde{\gamma}_{k} = \frac{\left| \mathbf{h}_k^H \mathbf{w}_k \right|^2}{\sum_{i \neq k} \left| \mathbf{h}_k^H \mathbf{w}_i \right|^2 + \sigma^2},
\end{equation}   
where $\mathbf{h}_k \in \mathbb{C}^{N \times 1}$ and $\mathbf{w}_k \in \mathbb{C}^{N \times 1}$ denotes the channel vector and beamformer for user $k$, respectively. Then, the general system utility maximization problem for SPDAs is given by 
\begin{subequations} \label{problem_SPDA}
    \begin{align}
       \max_{\mathbf{W}} \quad &U(\tilde{\boldsymbol{\gamma}}) \\
        \mathrm{s.t.} \quad & \sum_{k=1}^K \|\mathbf{w}_k\|^2 \le P,
    \end{align}
\end{subequations}
where $\mathbf{W} = [\mathbf{w}_1,\dots,\mathbf{w}_K]$ and $\tilde{\boldsymbol{\gamma}} = [\tilde{\gamma}_1,\dots,\tilde{\gamma}_K]^T$. We also define the channel matrix as $\mathbf{H} = [\mathbf{h}_1,\dots,\mathbf{h}_K]$. In Table \ref{table_comparison}, we compare the MRT, ZF, and MMSE designs and the optimal structure for SPDA beamforming, which have been extensively discussed in the literature \cite{6832894}, to their counterparts for CAPA beamforming derived in this paper. 

\begin{remark}
    \normalfont
    As shown in Table \ref{table_comparison}, the MRT, ZF, and MMSE designs, as well as the optimal structure, for both CAPA and SPDA beamforming share similar forms. Specifically, the beamformers for both CAPAs and SPDAs should be linear combinations of the channel response due to the reason discussed in \textbf{Remark \ref{remark_2}}. Moreover, the coefficients of these linear combinations are determined by the channel correlation matrix, which is $\mathbf{Q}$ for CAPAs and $\mathbf{H}^H \mathbf{H}$ for SPDAs.  
\end{remark}

\begin{remark} \label{remark_CAPA_gain}
    \normalfont
    \textcolor{black}{There are three main categories of benefit brought by using antenna arrays: 1) \emph{beamforming gain}, which is the ability to focus radiated energy toward users, 2) \emph{multiplexing gain}, which is the ability to transmit multiple independent data streams simultaneously, and 3) \emph{diversity gain}, which is the ability to improvement in link reliability by exploiting independent fading paths to reduce the impact of channel fading. As shown in the literature \cite{ouyang2024diversity}, when comparing CAPA with SPDA using half-wavelength or smaller antenna spacing, the primary performance improvement comes from enhanced beamforming gain. However, when comparing CAPA with SPDA using larger antenna spacing, CAPA also provides significant improvements in multiplexing and diversity gain.}
\end{remark}

\section{Numerical Results} \label{sec:results}
In this section, numerical results obtained through Monte Carlo simulations are provided to verify the effectiveness of the proposed optimal and heuristic designs. The following simulation setup is exploited throughout our simulations unless otherwise specified. It is assumed that the transmit CAPA is deployed within the $x$-$y$ plane and centered at the origin of the coordinate system, where 
\begin{equation}
  \mathcal{S} = \left\{ \mathbf{s} = [s_x, s_y, 0]^T \Big| |s_x| \le \frac{L_x}{2}, |s_y| \le \frac{L_y}{2} \right\},
\end{equation}
with $L_x = L_y = \sqrt{A_{\mathrm{T}}}$ and $A_{\mathrm{T}} = 0.1 \text{ m}^2$. There are $K = 4$ communication users randomly located within the following region 
\begin{equation}
  \mathcal{U} = \left\{ \mathbf{r} = [r_x, r_y, r_z]^T \Bigg| \begin{matrix}
    &|r_x| \le U_x, |r_y| \le U_y,\\
    &U_{z, \min} \le r_z \le U_{z, \max}
  \end{matrix} \right\},
\end{equation} 
with $U_x = U_y = 5$ m, $U_{z, \min} = 15$ m, and $U_{z, \max} = 30$ m. The free-space line-of-sight electromagnetic model is adopted to characterize the channel response $h_k(\mathbf{s})$ for each user $k$, which is given by \cite{wang2024beamforming} 
\begin{align}
  h_k^*(\mathbf{s}) = - &\frac{j \eta e^{- j \frac{2\pi}{\lambda} \|\mathbf{r}_k - \mathbf{s}\|}}{2 \lambda \|\mathbf{r}_k - \mathbf{s}\|} \nonumber \\
  &\times \hat{\mathbf{u}}_k^T \left( \mathbf{I}_3 - \frac{(\mathbf{r}_k-\mathbf{s})(\mathbf{r}_k - \mathbf{s})^T}{ \|\mathbf{r}_k - \mathbf{s}\|^2} \right) \hat{\mathbf{u}}_{\mathrm{T}},
\end{align} 
where $\eta = 120\pi$ $\Omega$ denotes the free-space intrinsic impedance, $\lambda$ denotes the signal wavelength, $\mathbf{r}_k \in \mathcal{U}$ denotes the location of user $k$, and $\hat{\mathbf{u}}_k \in \mathbb{R}^{3 \times 1}$ and $\hat{\mathbf{u}}_{\mathrm{T}} \in \mathbb{R}^{3 \times 1}$ denote the unit polarization direction of the receive antenna at user $k$ and transmit CAPA, respectively. Without loss of generality, we set $\hat{\mathbf{u}}_{\mathrm{T}} = \hat{\mathbf{u}}_k = [0,1,0]^T, \forall k$. The signal frequency is set to $2.4$ GHz. The transmit power and noise power are set to $P = 1$ mA$^2$ and \textcolor{black}{$\sigma^2 = 5.6 \times 10^{-3}$ $\text{V}^2/\text{m}^2$}, respectively. All integrals in the simulation are calculated using the Gauss-Legendre quadrature \cite{olver2010nist} with $20$ samples. The achievable sum rate is considered as the objective function for example:
\begin{equation}
  U(\boldsymbol{\gamma}) = \sum_{k=1}^K \log_2 \left(1 + \gamma_k\right).
\end{equation} 
With this objective function, the closed-form power allocation scheme in \cite[Theorem 3.16]{bjornson2013optimal} is exploited for the heuristic MRT, ZF, and MMSE designs.

\subsection{Simulation Benchmarks}
\textcolor{black}{For comparison, the conventional SPDA is used as a benchmark. The corresponding optimization problem for SPDA has been outlined in Section \ref{section:SPDA}. In this section, we introduce how to model the channel vectors $\mathbf{h}_k$ in a manner consistent with the CAPA framework, following the method in \cite{9906802} and \cite{zhang2023pattern}. Specifically, we consider an SPDA where the surface $\mathcal{S}$ is occupied with discrete antennas with $A_d = \frac{\lambda^2}{4\pi}$ effective area and $d = \frac{\lambda}{2}$ spacing. Therefore, there are totally $ N_d = \lceil \frac{2 L_x}{\lambda}\rceil \times \lceil \frac{2 L_y}{\lambda} \rceil$ antennas. The location of the $(n_x, n_y)$-th discrete antenna is given by
\begin{equation}
  \bar{\mathbf{s}}_{n_x, n_y} = \left[ (n_x-1)d - \frac{L_x}{2}, (n_y-1)d - \frac{L_y}{2}, 0 \right]^T.
\end{equation} 
Let $\mathcal{S}_{n_x, n_y}$ denote the effective area of the $(n_x, n_y)$-th discrete antenna. Then, the channel between this antenna and user $k$ can be calculated as \cite{9906802, zhang2023pattern}
\begin{equation} \label{SPDA_channel_model}
  h_{k, n_x, n_y} = \frac{1}{\sqrt{A_d}} \int_{\mathcal{S}_{n_x, n_y}}  h_k(\mathbf{s}) d \mathbf{s} \approx \sqrt{A}_d  h_k(\bar{\mathbf{s}}_{n_x, n_y}).
\end{equation}
The SPDA channel vector $\mathbf{h}_k$ in \eqref{SPDA_SINR} is then obtained by stacking all $h_{k, n_x, n_y}$ into a single vector. It is worth noting that with the channel model in \eqref{SPDA_channel_model}, the SPDA power constraint $\sum_{k=1}^K \|\mathbf{w}_k\|^2 \le P$ guarantees a fair signal and noise power setting in alignment with the CAPA power constraint $\sum_{k=1}^K \int_{\mathcal{S}} |w_k(\mathbf{s})|^2 d \mathbf{s} \le P$ \cite{9906802, zhang2023pattern}.    
Then, the optimal SPDA beamforming design for maximizing the achieved sum rate can be obtained using the monotonic optimization method described in \cite{zhang2013monotonic} and \cite{bjornson2013optimal}.}

\begin{figure}[t!]
  \centering
  \includegraphics[width=0.48\textwidth]{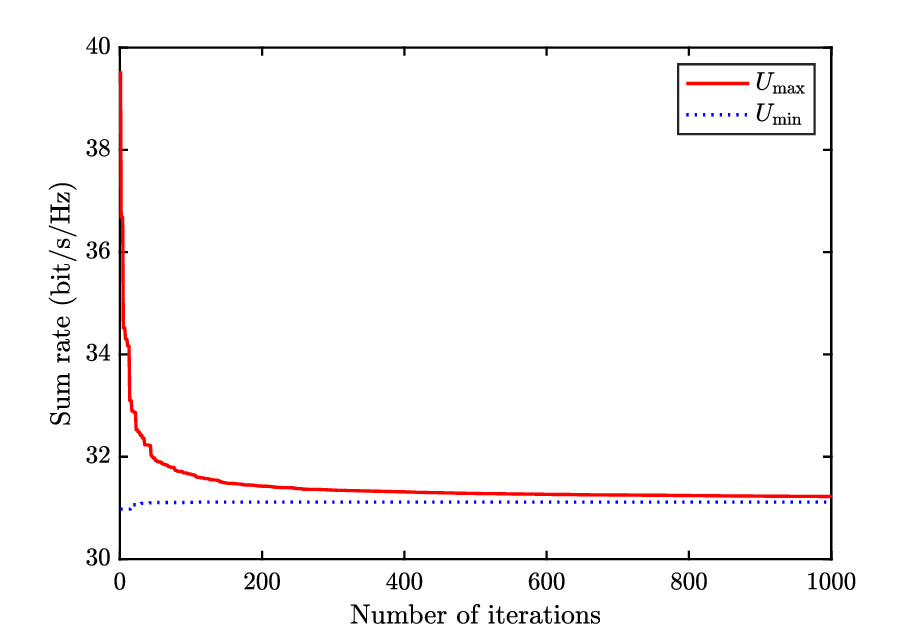}
  \caption{Convergence behavior of \textbf{Algorithm \ref{alg:optimal}}.}
  \label{fig_convergence}
  \vspace{-0.5cm}
\end{figure} 

\begin{figure}[t!]
  \centering
  \includegraphics[width=0.48\textwidth]{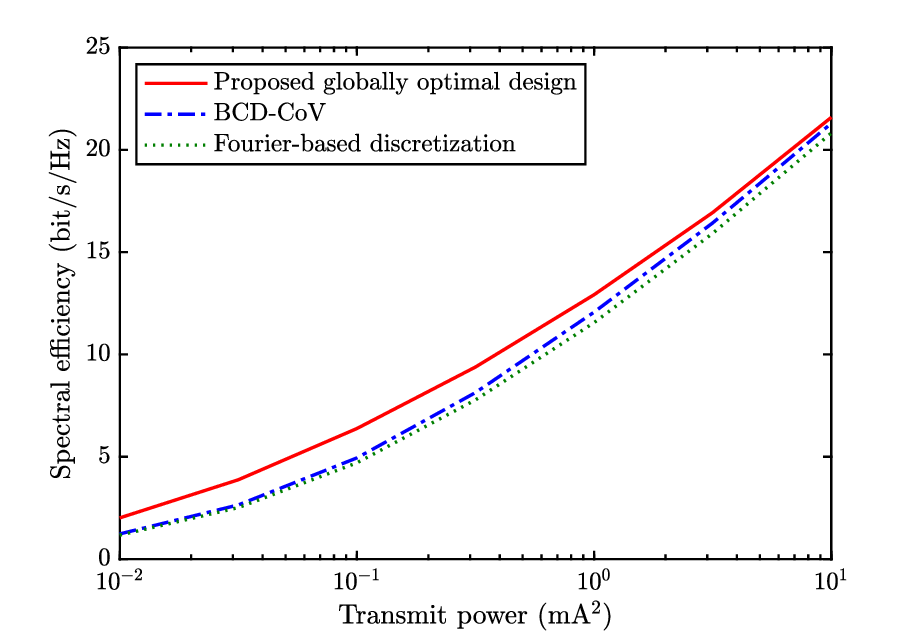}
  \caption{\textcolor{black}{Comparison with existing suboptimal methods.}}
  \label{fig_comparison}
\end{figure}

\subsection{Performance of the Proposed Globally Optimal Design}
We first evaluate the performance of the globally optimal design obtained using the proposed polyblock outer approximation algorithm.
Fig. \ref{fig_convergence} illustrates the convergence behavior of the proposed algorithm, as outlined in \textbf{Algorithm \ref{alg:optimal}}, for determining the optimal CAPA beamformer in a scenario with $K = 3$ users. It can be observed that the gap between the upper bound $U_{\max}$ and the lower bound $U_{\min}$ of the globally optimal objective value gradually decreases as the iterations progress. However, approximately $1000$ iterations are required for the gap to approach zero, indicating the high cost for finding the globally optimal solution. \textcolor{black}{Furthermore, Fig. \ref{fig_comparison} shows the performance gain of the proposed globally optimal design over the existing suboptimal methods, including the BCD-CoV method \cite{wang2024beamforming} and the Fourier-based discretization method~\cite{zhang2023pattern}, which confirms the effectiveness of the proposed design.}

\subsection{Impact of Transmit Power}
\begin{figure}[t!]
  \centering
  \includegraphics[width=0.48\textwidth]{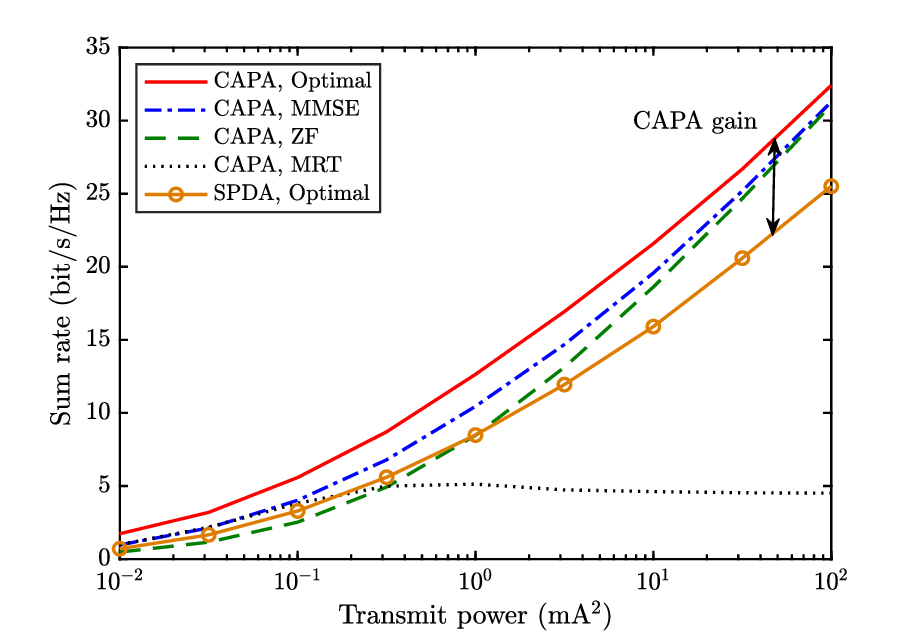}
  \caption{Sum rate versus transmit power.}
  \label{fig_transmit_power}
  \vspace{-0.5cm}
\end{figure}

Fig. \ref{fig_transmit_power} shows the sum rate achieved by different methods as a function of transmit power $P$. It is observed that the sum rate achieved by all methods increases with transmit power, except for the MRT method. This behavior occurs because the MRT method focuses solely on maximizing the power of the desired signal, without considering inter-user interference, which becomes significantly large in the high-SNR regime. Additionally, the proposed globally optimal CAPA design demonstrates a substantial performance gain over the globally optimal SPDA design. For instance, the proposed design increases the sum rate by $30\%$ from $20.5$ to $26.7$ bit/s/Hz when $P = 10$ mA$^2$. Furthermore, the MMSE design achieves performance close to the optimal design across the entire SNR regime. This is expected because the MMSE design adheres to the optimal beamforming structure and effectively enhances the SLNR for each user. In the high-SNR regime, the ZF design exhibits nearly identical performance to the MMSE design, owing to its asymptotic optimality.

\begin{figure}[t!]
  \centering
  \includegraphics[width=0.48\textwidth]{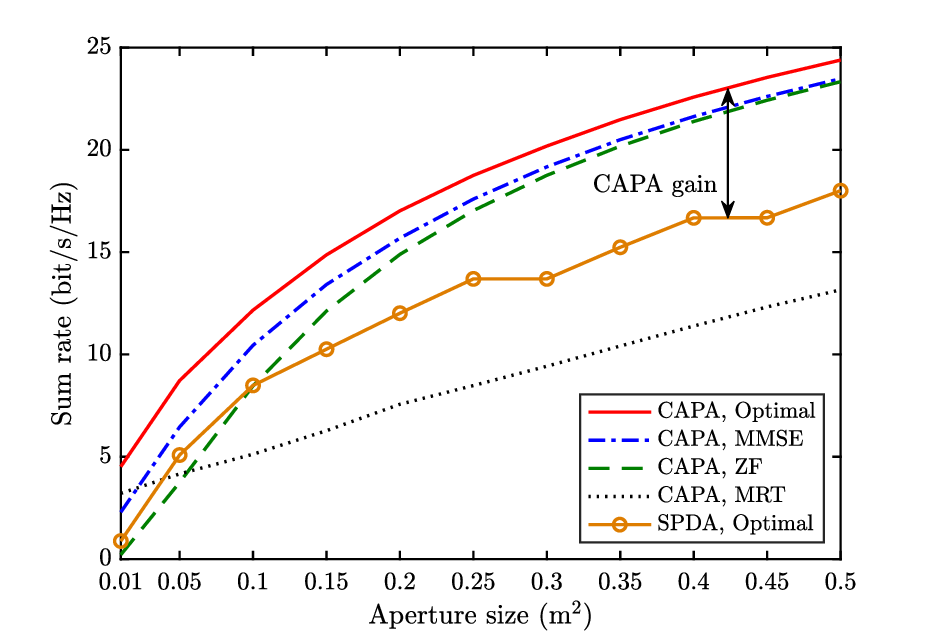}
  \caption{\textcolor{black}{Sum rate versus aperture size.}}
  \label{fig_aperture}
\end{figure}

\subsection{Impact of Aperture Size}
Fig. \ref{fig_aperture} shows the sum rate achieved by different methods as a function of the aperture size $A_{\mathrm{T}}$ of the transmit CAPA. As expected, the proposed globally optimal design for CAPA achieves the best performance across the considered range of aperture sizes and demonstrates a significant performance gain over the optimal design for SPDA. For example, a performance improvement of $37\%$ is observed when $A_{\mathrm{T}} = 0.25 \, \text{m}^2$. The MMSE design consistently delivers performance close to the optimal solution, owing to its ability to balance the power of desired signals with inter-user interference. Furthermore, the ZF and MRT designs exhibit near-optimal performance in cases of large and small apertures, respectively. This behavior is expected: with smaller apertures, the channels of different users are more likely to align with each other, causing the beamformer direction for eliminating inter-user interference to become nearly orthogonal to the channel directions. Conversely, as the aperture size increases, the channels between different users become closer to orthogonal, making the ZF design more effective.

\begin{figure}[t!]
  \centering
  \includegraphics[width=0.48\textwidth]{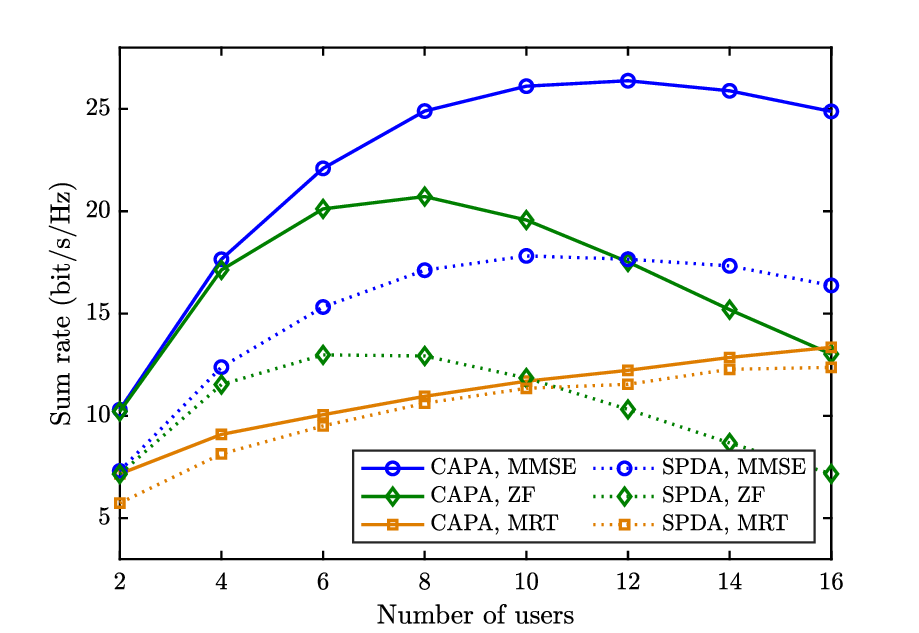}
  \caption{Sum rate versus the number of users.}
  \label{fig_user_sum}
  \vspace{-0.5cm}
\end{figure} 

\begin{figure}[t!]
  \centering
  \includegraphics[width=0.48\textwidth]{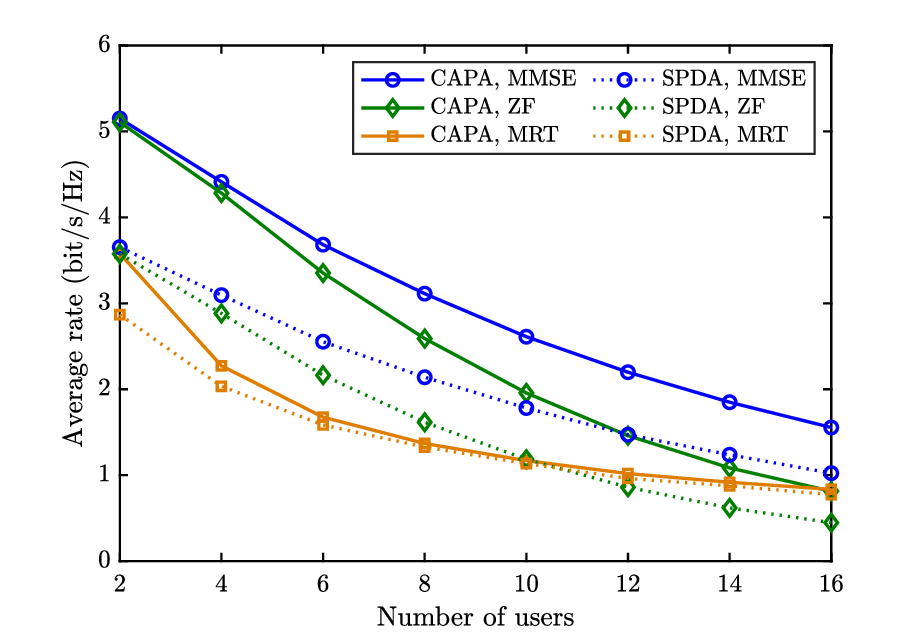}
  \caption{Average rate versus the number of users.}
  \label{fig_user_average}
\end{figure} 

\subsection{Impact of Number of Users}

Fig. \ref{fig_user_sum} illustrates the sum rate achieved by different methods as a function of the number of users. Here, we set $A_{\mathrm{T}} = 0.25 \, \text{m}^2$ and $P = K P_u$ with $P_u = 0.25 \, \text{mA}^2$.    It is observed that the sum rate achieved by both the MMSE and ZF designs initially increases and then decreases as the number of users grows. The peak sum rate for the MMSE design occurs at a higher number of users compared to the ZF design, indicating the ability of the MMSE design to better exploit the spatial DoFs provided by the continuous aperture to support more communication users. Additionally, the ZF design outperforms the MRT design when the number of users is small, while the opposite occurs as the number of users increases. This indicates that the spatial DoFs offered by a fixed CAPA aperture can support only a limited number of users for completely eliminating inter-user interference, even though the number of antennas can theoretically be considered infinite. Finally, when the MRT design is adopted, CAPA provides only marginal performance gains over SPDA, suggesting that the MRT design fails to efficiently utilize the spatial DoFs. Fig. \ref{fig_user_average} further shows the impact of the number of users on the average rate, defined as $\frac{1}{K} \sum_{k=1}^K \log_2(1 + \gamma_k)$. 
It can be observed that the average rate achieved by all methods decreases monotonically as the number of users increases due to the stronger inter-user interference.

\begin{figure}[t!]
  \centering
  \includegraphics[width=0.48\textwidth]{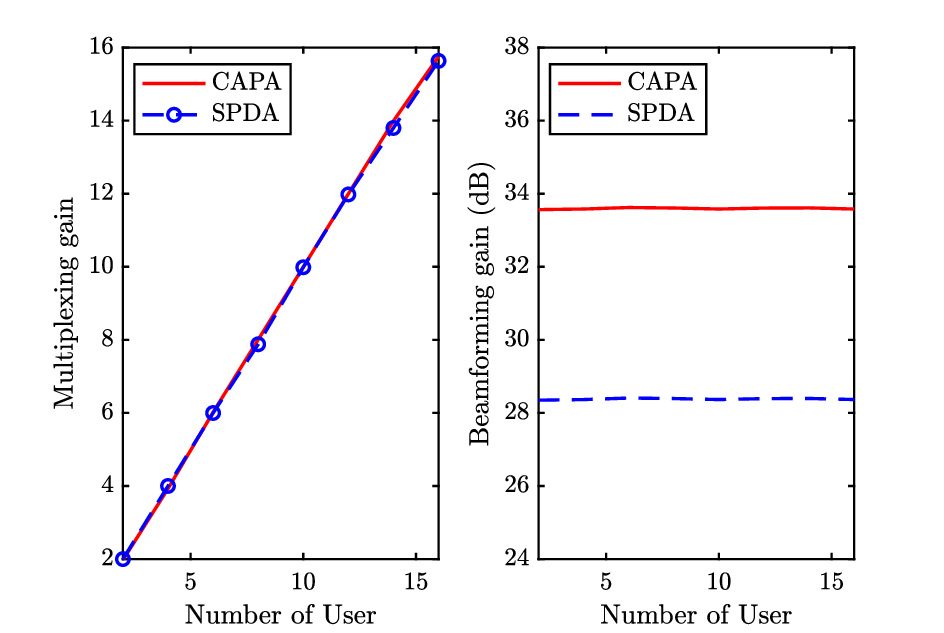}
  \caption{\textcolor{black}{Multiplexing gain (left) and beamforming gain (right).}}
  \label{fig_capa_gain}
\end{figure} 

\subsection{Multiplexing and Beamforming Gain}

\textcolor{black}{Fig. \ref{fig_capa_gain} provides further insights into the performance gains achieved by CAPA over SPDA. Since our simulation considers a line-of-sight channel, we primarily focus on multiplexing gain and beamforming gain, as discussed in \textbf{Remark \ref{remark_CAPA_gain}}. Specifically, the multiplexing gain is defined as 
\begin{equation}
  \lim_{P \rightarrow +\infty} \frac{\sum_{k=1}^K \log_2 \left(1 + \gamma_k\right)}{\log_2(P)},
\end{equation}
which represents the asymptotic slope of the spectral efficiency with respect to the logarithm of the transmit power. For the beamforming gain, we evaluate the following averaged values over all users for CAPA and SPDA, respectively:
\begin{align}
  G_{\mathrm{CAPA}} &= \frac{1}{K} \sum_{k=1}^K \int_{\mathrm{S}} |h_k(\mathbf{s})|^2  d \mathbf{s}, \\
 G_{\mathrm{SPDA}} &= \frac{1}{K} \sum_{k=1}^K \|\mathbf{h}_k\|^2.
\end{align}
As can be observed from Fig. \ref{fig_capa_gain}, CAPA exhibits the same multiplexing gain as SPDA but significantly improves the beamforming gain over SPDA, which is consistent with our discussion in \textbf{Remark \ref{remark_CAPA_gain}}.
}

\section{Conclusions} \label{sec:conclusion}

This paper has proposed a globally optimal design for CAPA multi-user transmit beamforming. Specifically, the optimal structure of the CAPA transmit beamforming was derived as a linear combination of the continuous channel responses, leveraging Lagrangian duality and the calculus of variations. Based on this structure, we proposed a monotonic optimization method to achieve the globally optimal transmit beamforming design for CAPA. We further derived the low-complexity heuristic MRT, ZF, and MMSE designs for CAPA transmit beamforming, which exhibit similar forms to their SPDA counterparts. The designs and insights provided in this paper offer valuable benchmarks for future CAPA system developments. \textcolor{black}{Additionally, accurate channel knowledge is essential to fully exploit the performance of all transmit beamforming designs proposed in this paper. Compressive sensing and parametric estimation methods, which leverage the inherent structure of the channel, is promising for addressing the continuous channel estimation challenge in CAPA. This remains an important direction for future research.}

\begin{appendices}

    \section{Proof of Lemma \ref{Lemma_CoV}} \label{Lemma_CoV_proof}
    Note that \eqref{fundamental_lemma_condition} must be satisfied for all possible smooth functions $\eta_k(\mathbf{s}), \forall k$. Therefore, it is sufficient to consider a special case. 
    Let us consider a special case where the functions $\eta_k(\mathbf{s})$ are given by 
    \begin{equation}
        \eta_k(\mathbf{s}) = e_k(\mathbf{s}) w_k(\mathbf{s}), \forall k, 
    \end{equation} 
    where $e_k(\mathbf{s})$ is an arbitrary function satisfying $e_k(\mathbf{s})> 0, \forall \mathbf{s} \in \mathcal{S}$ and $e_k(\mathbf{s}) = 0, \forall \mathbf{s} \in \partial \mathcal{S}$. Under this circumstance, the condition \eqref{fundamental_lemma_condition} becomes
    \begin{equation} 
        \sum_{k=1}^K \Re \left\{ \int_{\mathcal{S}} |w_k (\mathbf{s})|^2 e (\mathbf{s}) d \mathbf{s} \right\} = 0.
    \end{equation}
    It is readily to show that $w_k(\mathbf{s}) = 0, \forall k,$ must be satisfied under the above condition. This completes the proof.

    \section{Proof of Lemma \ref{lemma_inverse}} \label{lemma_inverse_proof}
    
    Substituting \eqref{general_form} and \eqref{general_form_inverse} into \eqref{left_inverse} yields
    \begin{align} \label{inverse_multiply}
        &\int_{\mathcal{S}} G^{-1} (\mathbf{r}, \mathbf{s}) G(\mathbf{s}, \mathbf{z}) d \mathbf{s} \nonumber \\
        &= \delta(\mathbf{r} - \mathbf{z}) + \sum_{i=1}^K \rho_i h_i(\mathbf{r}) h_i^*(\mathbf{z}) - \sum_{k=1}^K \sum_{i=1}^K \rho_i d_{k,i} h_k(\mathbf{r}) h_i^*(\mathbf{z}) \nonumber \\
        & \hspace{0.5cm}-  \sum_{k=1}^K \sum_{i=1}^K \sum_{j=1}^K \rho_i \rho_j d_{k,j} h_k (\mathbf{r}) h_i^*(\mathbf{z}) \int_{\mathcal{S}} h_i(\mathbf{s}) h_j^*(\mathbf{s}) d \mathbf{s} \nonumber \\
        & =\delta(\mathbf{r} - \mathbf{z}) + \sum_{k=1}^K \sum_{i=1}^K \rho_i \alpha_{k,i} h_k(\mathbf{r}) h_i^*(\mathbf{z}),
    \end{align}
    where
    \begin{align}
        \alpha_{k,i} = &u_{k,i}  - d_{k,i} - \sum_{j=1}^K \rho_j d_{k,j} \int_{\mathcal{S}} h_i(\mathbf{s}) h_j^*(\mathbf{s}) d \mathbf{s} \nonumber \\
        = &u_{k,i}  - d_{k,i} - \sum_{j=1}^K \rho_j d_{k,j} q_{j,i},
    \end{align} 
    where $u_{k,i} = 1$ if $k = i$; otherwise, $u_{k,i} = 0$.  
    Note that $\alpha_{k,i}$ is the entry of the following matrix in the $k$-th row and $j$-th column:
    \begin{align}
        \mathbf{I}_K - \mathbf{D} - \mathbf{D} \mathbf{\Lambda} \mathbf{Q} = &\mathbf{I}_K - (\mathbf{I} - \mathbf{\Lambda} \mathbf{Q})^{-1} - (\mathbf{I}_K - \mathbf{\Lambda} \mathbf{Q})^{-1} \mathbf{\Lambda} \mathbf{Q} \nonumber \\
        = & \mathbf{I}_K - (\mathbf{I}_K - \mathbf{\Lambda} \mathbf{Q})^{-1} (\mathbf{I}_K - \mathbf{\Lambda} \mathbf{Q}) \nonumber \\
        = &\mathbf{I}_K - \mathbf{I}_K = \mathbf{0}.
    \end{align} 
    Based on the above result, we have 
    \begin{equation} \label{zero_alpha}
        \alpha_{k,i} = 0, \forall k, i.
    \end{equation}
    Substituting \eqref{zero_alpha} into \eqref{inverse_multiply} yields
    \begin{equation}
        \int_{\mathcal{S}} G^{-1} (\mathbf{r}, \mathbf{s}) G(\mathbf{s}, \mathbf{z}) d \mathbf{s} = \delta(\mathbf{r} - \mathbf{z}).
    \end{equation}
    This proof is thus completed.
    
    \section{Proof of Lemma \ref{lemma_power}} \label{lemma_power_proof}
    Let $\tilde{\mathbf{w}}(\mathbf{s}) = [\tilde{w}_1(\mathbf{s}), \dots, \tilde{w}_K(\mathbf{s})]$ denote a feasible solution to problem \eqref{problem_1} that does not satisfy the power constraint with equality, i.e., $\sum_{k=1}^K \int_{\mathcal{S}} |\tilde{w}_k(\mathbf{s})|^2 \, d\mathbf{s} < P$. Define the following scaling factor:
    \begin{align}
        \zeta = \frac{P}{\sum_{k=1}^K \int_{\mathcal{S}} |\tilde{w}_k(\mathbf{s})|^2 \, d\mathbf{s}},
    \end{align}
    which satisfies $\zeta > 1$. It is straightforward to show that the scaled solution $\breve{\mathbf{w}}(\mathbf{s}) = \sqrt{\zeta} \, \tilde{\mathbf{w}}(\mathbf{s})$ meets the power constraint with equality. The SINR achieved by the solution $\breve{\mathbf{w}}(\mathbf{s})$ for user $k$ is given by
    \begin{align}
        \breve{\gamma}_k =& \frac{\zeta \left| \int_{\mathcal{S}} h_k^*(\mathbf{s}) \tilde{w}_k(\mathbf{s}) \, d\mathbf{s} \right|^2}{\sum_{i \neq k} \zeta \left| \int_{\mathcal{S}} h_k^*(\mathbf{s}) \tilde{w}_i(\mathbf{s}) \, d\mathbf{s} \right|^2 + \sigma^2} \nonumber \\
        >& \frac{\left| \int_{\mathcal{S}} h_k^*(\mathbf{s}) \tilde{w}_k(\mathbf{s}) \, d\mathbf{s} \right|^2}{\sum_{i \neq k} \left| \int_{\mathcal{S}} h_k^*(\mathbf{s}) \tilde{w}_i(\mathbf{s}) \, d\mathbf{s} \right|^2 + \sigma^2} = \tilde{\gamma}_k,
    \end{align}
    where the last inequality follows from $\zeta > 1$ and $\tilde{\gamma}_k$ is the SINR achieved by the original solution $\tilde{\mathbf{w}}(\mathbf{s})$ for user $k$. Since the utility function $U(\cdot)$ is strictly monotonically increasing in each $\gamma_k$, the scaled solution $\breve{\mathbf{w}}(\mathbf{s})$ achieves a higher objective value than the original solution $\tilde{\mathbf{w}}(\mathbf{s})$. Therefore, the optimal solution must satisfy the power constraint with equality. This proof is thus completed.

    \section{Proof of Theorem \ref{theorem_structure}} \label{theorem_structure_proof}
    
    We exploit the duality of optimization theory and the calculus of variations to prove \textbf{Theorem \ref{theorem_structure}}. First, we note that problem \eqref{problem_2} is convex and satisfies strong duality. This can be demonstrated by considering $h_k(\mathbf{s})$ and $w_k(\mathbf{s}), \forall k, $ as infinitesimal vectors, where $h_k(\tilde{\mathbf{s}})$ and $w_k(\tilde{\mathbf{s}})$ for a particular $\tilde{\mathbf{s}} \in \mathcal{S}$ represent individual entries of these vectors. Then, by applying the results from \cite{1561584} and \cite{6832894}, the convexity and strong duality of the problem can be established.
    
    To streamline the proof, we rewrite the constraint in problem \eqref{problem_2} into the following equivalent form:
    \begin{align}
        &\frac{1}{\eta_k \sigma^2} \left| \int_{\mathcal{S}} h_k^*(\mathbf{s}) w_k(\mathbf{s}) d \mathbf{s} \right|^2 \nonumber \\ &\ge \frac{1}{\sigma^2} \sum_{i\neq k} \left| \int_{\mathcal{S}} h_k^*(\mathbf{s}) w_i(\mathbf{s}) d \mathbf{s} \right|^2 + 1, \forall k.
    \end{align}
    Therefore, the Lagrangian function of problem \eqref{problem_2} is
    \begin{align}
        &L\left(\mathbf{w}(\mathbf{s}), \boldsymbol{\lambda}\right) \nonumber \\ \hspace{0.2cm} &= \sum_{k=1}^K \int_{\mathcal{S}} |w_k(\mathbf{s})|^2 d \mathbf{s} + \sum_{k=1}^K \lambda_k  \Bigg(\frac{1}{\sigma^2} \sum_{i\neq k} \left| \int_{\mathcal{S}} h_k^*(\mathbf{s}) w_i(\mathbf{s}) d \mathbf{s} \right|^2 \nonumber\\
        &\hspace{0.5cm}+ 1 - \frac{1}{\eta_k \sigma^2} \left| \int_{\mathcal{S}} h_k^*(\mathbf{s}) w_k(\mathbf{s}) d \mathbf{s} \right|^2  \Bigg) \nonumber \\
        & \hspace{0.2cm} = \sum_{k=1}^K \Bigg( \int_{\mathcal{S}} |w_k(\mathbf{s})|^2 d \mathbf{s} + \sum_{i\neq k} \frac{\lambda_i}{\sigma^2} \left| \int_{\mathcal{S}} h_i^*(\mathbf{s}) w_k(\mathbf{s}) d \mathbf{s} \right|^2 \nonumber \\
        & \hspace{0.5cm}- \frac{\lambda_k}{\eta_k \sigma^2} \left| \int_{\mathcal{S}} h_k^*(\mathbf{s}) w_k(\mathbf{s}) d \mathbf{s} \right|^2 + \lambda_k \Bigg),
    \end{align}
    where $\boldsymbol{\lambda} = [\lambda_1,\dots,\lambda_K]^T$ and $\lambda_k \ge 0$ is the Lagrange multiplier. 
    It is easy to show that the Lagrange dual function of problem \eqref{problem_2} is
    \begin{align}
        D(\boldsymbol{\lambda}) = \inf_{\mathbf{w}(\mathbf{s})} L\left(\mathbf{w}(\mathbf{s}), \boldsymbol{\lambda}\right) = \begin{cases}
            \sum_{k=1}^K \lambda_k & \text{if } t_k \ge 0, \forall k, \\
            -\infty  & \text{otherwise},
        \end{cases}
    \end{align}
    where $t_k = \int_{\mathcal{S}} |w_k(\mathbf{s})|^2 d \mathbf{s} + \sum_{i\neq k} \frac{\lambda_i}{\sigma^2} \left| \int_{\mathcal{S}} h_i^*(\mathbf{s}) w_k(\mathbf{s}) d \mathbf{s} \right|^2 \nonumber - \frac{\lambda_k}{\eta_k \sigma^2} \left| \int_{\mathcal{S}} h_k^*(\mathbf{s}) w_k(\mathbf{s}) d \mathbf{s} \right|^2$. Therefore, the optimal value of the dual problem satisfies $\max_{\boldsymbol{\lambda}} D(\boldsymbol{\lambda}) = \sum_{k=1}^K \lambda_k$. According to the strong duality, problem \eqref{problem_2} and its dual problem have the same objective value at the optimum, i.e.,
    \begin{equation} \label{dual_gap}
        \sum_{k=1}^K \lambda_k = \sum_{k=1}^K \int_{\mathcal{S}} |w_k(\mathbf{s})|^2 d \mathbf{s} = P,
    \end{equation}
    where the last step is obtained using \textbf{Lemma \ref{lemma_power}}.
    
    Furthermore, based on KKT conditions, the condition for the optimal solution $\mathbf{w}(\mathbf{s})$ to problem \eqref{problem_2} is that $L\left(\mathbf{w}(\mathbf{s}), \boldsymbol{\lambda}\right)$ reaches its minimum. However, since $\mathbf{w}(\mathbf{s})$ is a function rather than a discrete variable, the conventional first-order derivative condition cannot be applied to find the minimum of $L\left(\mathbf{w}(\mathbf{s}), \boldsymbol{\lambda}\right)$. To address this, we apply the calculus of variations from functional analysis. Specifically, at the minimum of $L\left(\mathbf{w}(\mathbf{s}), \boldsymbol{\lambda}\right)$, for any $\epsilon \rightarrow 0$, we have 
    \begin{equation} \label{Cov_1}
        L\left(\mathbf{w}(\mathbf{s}), \boldsymbol{\lambda}\right) \le L\left(\mathbf{w}(\mathbf{s}) + \epsilon \boldsymbol{\eta}(\mathbf{s}), \boldsymbol{\lambda}\right) \triangleq \Phi(\epsilon),
    \end{equation}  
    where $\boldsymbol{\eta}(\mathbf{s}) = [\eta_1(\mathbf{s}),\dots,\eta_K(\mathbf{s})]^T$ with $\eta_k(\mathbf{s}), k =1,\dots,K,$ being any arbitrary smooth function defined on $\mathcal{S}$ that satisfies $\eta_k(\mathbf{s}) = 0, \forall \mathbf{s} \in \partial \mathcal{S}$. In particular, $\epsilon \boldsymbol{\eta}(\mathbf{s})$ is referred to as the variation of $\mathbf{w}(\mathbf{s})$. At the minimum of $L\left(\mathbf{w}(\mathbf{s}), \boldsymbol{\lambda}\right)$, function $\Phi(\epsilon)$ must also have its minimum at $\epsilon = 0$ and thus 
    \begin{equation} \label{CoV_condition}
        \left.\frac{d \Phi(\epsilon)}{d \epsilon}\right|_{\epsilon = 0} = 0.
    \end{equation}
    To exploit the above condition, we first give the explicit expression of $\Phi(\epsilon)$ in the following: 
    \begin{align} \label{explicit_Phi}
        \Phi(\epsilon) =& \sum_{k=1}^K 2 \epsilon \Re\bigg\{  \int_{\mathcal{S}} w_k(\mathbf{s}) \eta_k^*(\mathbf{s}) d \mathbf{s} \nonumber \\
        &+  \sum_{i\neq k} \frac{\lambda_i}{\sigma^2} \int_{\mathcal{S}} \int_{\mathcal{S}} h_i^*(\mathbf{z}) w_k(\mathbf{z}) h_i(\mathbf{s}) \eta_k^*(\mathbf{s}) d \mathbf{z} d \mathbf{s}  \nonumber \\
        & - \frac{\lambda_k}{\eta_k \sigma^2} \int_{\mathcal{S}} \int_{\mathcal{S}} h_k^*(\mathbf{z}) w_k(\mathbf{z}) h_k(\mathbf{s}) \eta_k^*(\mathbf{s}) d \mathbf{z} d \mathbf{s} \bigg\}  \nonumber \\
        &+ \Psi(\epsilon^2) + C,
    \end{align}
    where $\Psi(\epsilon^2)$ collects all terms related to $\epsilon^2$ and $C$ is a constant irrelevant to $\epsilon$. Substituting \eqref{explicit_Phi} into \eqref{CoV_condition} yields
    \begin{equation}
        \sum_{k=1}^K \Re \left\{ \int_{\mathcal{S}} v_k(\mathbf{s}) \eta_k^*(\mathbf{s}) d \mathbf{s} \right\} = 0,
    \end{equation}
    where
    \begin{align}
        v_k(\mathbf{s}) = &w_k(\mathbf{s}) + \sum_{i \neq k} \frac{\lambda_i}{\sigma^2} h_i(\mathbf{s}) \int_{\mathcal{S}} h_i^*(\mathbf{z}) w_k(\mathbf{z}) d \mathbf{z} \nonumber \\
        & - \frac{\lambda_k}{\eta_k \sigma^2} h_k(\mathbf{s}) \int_{\mathcal{S}} h_k^*(\mathbf{z}) w_k(\mathbf{z}) d \mathbf{z} \nonumber \\
        = &w_k(\mathbf{s}) + \sum_{i = 1}^K \frac{\lambda_i}{\sigma^2} h_i(\mathbf{s}) \int_{\mathcal{S}} h_i^*(\mathbf{z}) w_k(\mathbf{z}) d \mathbf{z} \nonumber \\
        & - \frac{\lambda_k}{\sigma^2} (1 + \frac{1}{\eta_k}) h_k(\mathbf{s}) \int_{\mathcal{S}} h_k^*(\mathbf{z}) w_k(\mathbf{z}) d \mathbf{z}.
    \end{align}
    Then, based on \textbf{Lemma \ref{Lemma_CoV}}, it must hold that $v_k(\mathbf{s}) = 0, \forall k$, leading to 
    \begin{align} \label{theorem_condition_1}
        &w_k(\mathbf{s}) + \sum_{i = 1}^K \frac{\lambda_i}{\sigma^2} h_i(\mathbf{s}) \int_{\mathcal{S}} h_i^*(\mathbf{z}) w_k(\mathbf{z}) d \mathbf{z} \nonumber \\
        & \hspace{1cm} = \underbrace{\frac{\lambda_k}{\sigma^2} (1 + \frac{1}{\eta_k}) \int_{\mathcal{S}} h_k^*(\mathbf{z}) w_k(\mathbf{z}) d \mathbf{z}}_{\triangleq \sqrt{p_k}} h_k(\mathbf{s})
    \end{align}
    Based on the definition of the function $\delta(\mathbf{s} - \mathbf{z})$ in \eqref{identify_function}, we have $w_k(\mathbf{s}) = \int_{\mathcal{S}} \delta(\mathbf{s} - \mathbf{z}) w(\mathbf{z}) d \mathbf{z}$. Based on this identity, \eqref{theorem_condition_1} can be rewritten as 
    \begin{align} \label{theorem_condition_2}
        \int_{\mathcal{S}} \left( \delta(\mathbf{s}- \mathbf{z}) + \sum_{i=1}^K \frac{\lambda_i}{\sigma^2} h_i (\mathbf{s}) h_i^*(\mathbf{z})  \right) w_k(\mathbf{z}) d \mathbf{z} = \sqrt{p_k} h_k(\mathbf{s}).
    \end{align}
    Defining 
    \begin{equation}
        G(\mathbf{s}, \mathbf{z}) = \delta(\mathbf{s}- \mathbf{z}) + \sum_{i=1}^K \frac{\lambda_i}{\sigma^2} h_i (\mathbf{s}) h_i^*(\mathbf{z}),
    \end{equation}
    \eqref{theorem_condition_2} can be simplified as 
    \begin{align} \label{theorem_condition_3}
        \int_{\mathcal{S}} G(\mathbf{s}, \mathbf{z}) w_k(\mathbf{z}) d \mathbf{z} = \sqrt{p_k} h_k(\mathbf{s}).
    \end{align}
    Based on \textbf{Lemma \ref{lemma_inverse}}, the inverse of $G(\mathbf{s}, \mathbf{z})$ is given by 
    \begin{equation}
        G^{-1}(\mathbf{r}, \mathbf{s}) = \delta(\mathbf{r} - \mathbf{s}) - \sum_{k=1}^K \sum_{i=1}^K \frac{\lambda_i}{\sigma^2} d_{k,i} h_k(\mathbf{r}) h_i^*(\mathbf{s}),
    \end{equation} 
    Here, $d_{k,i}$ is the entry of matrix $\mathbf{D} = (\mathbf{I}_K + \frac{1}{\sigma^2}\mathbf{\Lambda}\mathbf{Q})^{-1}$ in the $k$-th row and $i$-th column, where $\mathbf{\Lambda}$ is a diagonal matrix defined as 
    \begin{equation}
        \mathbf{\Lambda} = \mathrm{diag} \left( \lambda_1,\dots,\lambda_K \right),
    \end{equation}   
    and the entry of matrix $\mathbf{Q}$ in the $k$-th row and $i$-th column is given by 
    \begin{equation}
        q_{k,i} = [\mathbf{Q}]_{k,i} = \int_{\mathcal{S}} h_i(\mathbf{s}) h_k^*(\mathbf{s}) d \mathbf{s}.
    \end{equation}
    Then, multiplying both sides of \eqref{theorem_condition_3} by $G^{-1}(\mathbf{r}, \mathbf{s})$ and integrating over $d \mathbf{s}$, we have 
    \begin{align}
        \int_{\mathcal{S}} \int_{\mathcal{S}} G^{-1}(\mathbf{r}, \mathbf{s}) G(\mathbf{s}, \mathbf{z}) w_k(\mathbf{z}) d \mathbf{s} d \mathbf{z} \nonumber \\ = \sqrt{p_k} \int_{\mathcal{S}} G^{-1}(\mathbf{r}, \mathbf{s}) h_k(\mathbf{s}) d \mathbf{s} \nonumber
    \end{align}
    \begin{align} \label{scalar_optimal_structure}
        \Leftrightarrow  w_k(\mathbf{r}) = &\sqrt{p_k} \int_{\mathcal{S}} G^{-1}(\mathbf{r}, \mathbf{s}) h_k(\mathbf{s}) d \mathbf{s} \nonumber \\
        =&\sqrt{p_k} \left(h_k(\mathbf{r}) - \sum_{i=1}^K \sum_{j=1}^K \frac{\lambda_j}{\sigma^2} d_{i,j} q_{j,k} h_i(\mathbf{r}) \right).
    \end{align}
    To simplify the above expression, we define the following variables:
    \begin{subequations}
        \begin{align}
            &\mathbf{\Lambda} = \mathrm{diag}(\lambda_1,\dots,\lambda_K), \\
            &\mathbf{P} = \mathrm{diag}(p_1,\dots,p_K), \\
            &\mathbf{w}(\mathbf{r}) = [w_1(\mathbf{r}),\dots,w_K(\mathbf{r})] \in \mathbb{C}^{1 \times K}, \\
            &\mathbf{h}(\mathbf{r}) = [h_1(\mathbf{r}),\dots,h_K(\mathbf{r})] \in \mathbb{C}^{1 \times K}.
        \end{align}
    \end{subequations}
    Then, \eqref{scalar_optimal_structure} can be written in matrix form as 
    \begin{align} \label{matrix_optimal_structure}
        \mathbf{w}(\mathbf{r}) &= \mathbf{h}(\mathbf{r}) \left( \mathbf{I}_K - \frac{1}{\sigma^2}\mathbf{D} \mathbf{\Lambda} \mathbf{Q}  \right) \mathbf{P}^{\frac{1}{2}} \nonumber \\
        &= \mathbf{h}(\mathbf{r}) \left( \mathbf{I}_K - \left(\mathbf{I}_K + \frac{1}{\sigma^2}\mathbf{\Lambda} \mathbf{Q} \right)^{-1} \frac{1}{\sigma^2}\mathbf{\Lambda} \mathbf{Q}  \right) \mathbf{P}^{\frac{1}{2}}  \nonumber \\
        &= \mathbf{h}(\mathbf{r}) \left(\mathbf{I}_K + \frac{1}{\sigma^2}\mathbf{\Lambda} \mathbf{Q} \right)^{-1} \mathbf{P}^{\frac{1}{2}},
    \end{align}
    where the last step is derived from the Woodbury matrix identity. The proof is thus completed.
    
    \section{Proof of Theorem \ref{theorem_SLNR}} \label{theorem_SLNR_proof}
    
    The SLNR maximization problem for user $k$ under the equal power allocation is given by 
    \begin{subequations} \label{SLNR_problem_1}
        \begin{align}
           \max_{w_k(\mathbf{s})} \quad &\mathrm{SLNR}_k \\
            \mathrm{s.t.} \quad & \int_{\mathcal{S}} |w_k(\mathbf{s)}|^2 d \mathbf{s} = \frac{P}{K}.
        \end{align}
    \end{subequations}
    Given that $\int_{\mathcal{S}} |w_k(\mathbf{s})|^2 d \mathbf{s} = \frac{P}{K}$, the SLNR for user $k$ can be rewritten as 
    \begin{equation}
        \overline{\mathrm{SLNR}}_k = \frac{\left| \int_{\mathcal{S}} h_k^*(\mathbf{s}) w_k(\mathbf{s}) d \mathbf{s} \right|^2}{\chi(w_k(\mathbf{s})) - \left| \int_{\mathcal{S}} h_k^*(\mathbf{s}) w_k(\mathbf{s}) d \mathbf{s} \right|^2},
    \end{equation} 
    where
    \begin{equation}
        \chi(w_k(\mathbf{s})) = \sum_{i=1}^K \left| \int_{\mathcal{S}} h_i^*(\mathbf{s}) w_k(\mathbf{s}) d \mathbf{s} \right|^2 + \frac{K \sigma^2}{P} \int_{\mathcal{S}} |w_k(\mathbf{s)}|^2 d \mathbf{s}.
    \end{equation}
    Let us now consider the following unconstrained optimization problem:
    \begin{equation} \label{SLNR_problem_2}
        \max_{w_k(\mathbf{s})} \quad \overline{\mathrm{SLNR}}_k.
    \end{equation} 
    Let $w_k^{\star}(\mathbf{s})$ and $w_k^{\dagger}(\mathbf{s})$ denote the optimal solutions to problems \eqref{SLNR_problem_1} and \eqref{SLNR_problem_2}, respectively. It is easy to verify that $w_k^{\star}(\mathbf{s})$ and $w_k^{\dagger}(\mathbf{s})$ are related as follows:
    \begin{equation}
        w_k^{\star}(\mathbf{s}) = \sqrt{ \frac{P}{K \int_{\mathcal{S}} |w_k^{\dagger}(\mathbf{s)}|^2 d \mathbf{s}}} w_k^{\dagger}(\mathbf{s}),
    \end{equation}
    which implies that $w_k^{\star}(\mathbf{s})$ is a scaled version of $w_k^{\dagger}(\mathbf{s})$, meaning they share the same structure. Therefore, we will focus on finding the optimal $w_k^{\dagger}(\mathbf{s})$ for maximizing $\overline{\mathrm{SLNR}}_k$. 
    
    Note that maximizing $\overline{\mathrm{SLNR}}_k$ is equivalent to minimizing its inverse. Therefore, optimal solution to problem \eqref{SLNR_problem_2} can be obtained by
    \begin{align} \label{SLNR_problem_3}
        w_k^{\dagger}(\mathbf{s}) = & \argmin_{w_k(\mathbf{s}) \in \mathcal{W}_k } \, \left(\chi(w_k(\mathbf{s})) - \left| \int_{\mathcal{S}} h_k^*(\mathbf{s}) w_k(\mathbf{s}) d \mathbf{s} \right|^2 \right) \nonumber \\
       \overset{(a)}{=} &\argmin_{\scriptstyle w_k(\mathbf{s}) \in \mathcal{W}_k } \, \chi(w_k(\mathbf{s})) 
        \overset{(b)}{=}\argmin_{w_k(\mathbf{s}) \in \overline{\mathcal{W}}_k } \, \chi(w_k(\mathbf{s})),
    \end{align}
    where 
    \begin{align}
        \mathcal{W}_k =& \left\{ w_k(\mathbf{s}) :  \left| \int_{\mathcal{S}} h_k^*(\mathbf{s}) w_k(\mathbf{s}) d \mathbf{s} \right|^2 = 1 \right\}, \\
        \overline{\mathcal{W}}_k =& \left\{ w_k(\mathbf{s}) : \Re \left\{ \int_{\mathcal{S}} h_k^*(\mathbf{s}) w_k(\mathbf{s}) d \mathbf{s} \right\} = 1 \right\}.
    \end{align}
    More specifically, the step $(a)$ follows from the fact that $| \int_{\mathcal{S}} h_k^*(\mathbf{s}) w_k(\mathbf{s}) d \mathbf{s} |^2$ has a constant value of $1$. Step $(b)$ results from the observation that $w_k(\mathbf{s})$ and $w_k(\mathbf{s}) e^{j \theta_k}$ are equivalent in minimizing the objective function for any rotation phase $\theta_k$. Thus, we can always choose $\theta_k$ such that $\int_{\mathcal{S}} h_k^*(\mathbf{s}) w_k(\mathbf{s}) d \mathbf{s}$ is real-valued. The optimal form of $w_k^{\dagger}(\mathbf{s})$ can be obtained using KKT conditions.  Specifically, the Lagrangian function to problem \eqref{SLNR_problem_3} is given by 
    \begin{equation}
        \tilde{L}(w_k(\mathbf{s}), \tilde{\lambda}_k) = \chi(w_k(\mathbf{s})) - \tilde{\lambda} \left( \Re \left\{ \int_{\mathcal{S}} h_k^*(\mathbf{s}) w_k(\mathbf{s}) d \mathbf{s} \right\} - 1 \right),
    \end{equation}
    where $\tilde{\lambda}_k \ge 0$ is the Lagrange multiplier. The optimal $w_k^{\dagger}(\mathbf{s})$ must be a stationary point of $\tilde{L}(w_k(\mathbf{s}), \tilde{\lambda}_k)$, which can be determined using a similar calculus of variations method as described in \eqref{Cov_1}-\eqref{theorem_condition_1}, resulting in the following optimality condition:
    \begin{align} \label{SLNR_eq_1}
        \frac{K \sigma^2}{P} w_k^{\dagger}(\mathbf{s}) + \sum_{k=1}^K h_i(\mathbf{s}) \int_{\mathcal{S}} h_i^*(\mathbf{z}) w_k^{\dagger}(\mathbf{z}) d \mathbf{z}  = \frac{\tilde{\lambda}_k}{2} h_k(\mathbf{s}).
    \end{align}
    By using the identity $w_k^{\dagger}(\mathbf{s}) = \int_{\mathcal{S}} \delta(\mathbf{s} - \mathbf{z}) w^{\dagger}(\mathbf{z}) d \mathbf{z}$, \eqref{SLNR_eq_1} can be reformulated as 
    \begin{equation} \label{SLNR_eq_2}
        \int_{\mathcal{S}} \tilde{G}(\mathbf{s}, \mathbf{z}) w_k^{\dagger}(\mathbf{z}) = \frac{\tilde{\lambda}_k P}{2 K \sigma^2} h_k(\mathbf{s}),
    \end{equation}
    where 
    \begin{equation}
        \tilde{G}(\mathbf{s}, \mathbf{z}) =  \delta(\mathbf{s} - \mathbf{z}) + \frac{P}{K \sigma^2} \sum_{k=1}^K h_i(\mathbf{s}) h_i^*(\mathbf{z}).
    \end{equation}
    Based on \textbf{Lemma \ref{lemma_inverse}}, the inverse of $\tilde{G}(\mathbf{s}, \mathbf{z})$ is given by
    \begin{equation} \label{SLNR_eq_3}
        \tilde{G}^{-1} (\mathbf{r}, \mathbf{s}) = \delta(\mathbf{r} - \mathbf{s}) - \frac{P}{K \sigma^2} \sum_{k=1}^K \sum_{i=1}^K  \tilde{d}_{k,i} h_k(\mathbf{r}) h_i^*(\mathbf{s}),
    \end{equation}
    where $\tilde{d}_{k,i}$ is the entry in the $k$-th row and $i$-th column of matrix $\tilde{\mathbf{D}} = (\mathbf{I}_K + \frac{P}{K \sigma^2} \mathbf{Q})^{-1}$.  Then, multiplying both sides of \eqref{SLNR_eq_2} by $\tilde{G}^{-1}(\mathbf{r}, \mathbf{s})$ and integrating over $d \mathbf{s}$, we have 
    \begin{align} \label{SLNR_eq_4}
         w_k^{\dagger}(\mathbf{r}) = \frac{\tilde{\lambda}_k P}{2 K \sigma^2} \int_{\mathcal{S}} \tilde{G}^{-1}(\mathbf{r}, \mathbf{s}) h_k(\mathbf{s}).
    \end{align}
    By substituting \eqref{SLNR_eq_3} into \eqref{SLNR_eq_4} and following a similar process as in \eqref{scalar_optimal_structure}-\eqref{matrix_optimal_structure}, the optimal beamforming vector $\mathbf{w}^{\dagger}(\mathbf{r}) = [w_1^{\dagger}(\mathbf{r}),\dots,w_K^{\dagger}(\mathbf{r})]$ can be derived as 
    \begin{equation} \label{SLNR_eq_5}
        \mathbf{w}^{\dagger}(\mathbf{r}) = \mathbf{h}(\mathbf{r}) \left( \mathbf{I}_K + \frac{P}{K \sigma^2} \mathbf{Q} \right)^{-1} \tilde{\mathbf{P}}^{\frac{1}{2}},
    \end{equation}
    where $\tilde{\mathbf{P}} = \mathrm{diag}(\tilde{p}_1,\dots,\tilde{p}_K)$ with $p_k = \left(\frac{\tilde{\lambda}_k P}{2 K \sigma^2} \right)^2$. It can be observed that \eqref{SLNR_eq_5} has the same form as \eqref{MMSE_deaign}. The proof is thus completed.
    
\end{appendices}

\balance
\bibliographystyle{IEEEtran}
\bibliography{reference/mybib}

\end{document}